\documentstyle[11pt,aaspp4]{article}

\newcommand{\etal}{{\it et al.\/}}
\newcommand{\kms}{\hbox{km~s$^{-1}$}}
\newcommand{\ha}{\hbox{H$\alpha$}}
\newcommand{\hb}{\hbox{H$\beta$}}
\newcommand{\hi}{\hbox{{\ion{H}{1}}}}
\newcommand{\hii}{\hbox{{\ion{H}{2}}}}
\newcommand{\nii}{\hbox{[{\ion{N}{2}}]}}
\newcommand{\sii}{\hbox{[{\ion{S}{2}}]}}
\newcommand{\oiii}{\hbox{[{\ion{O}{3}}]}}
\newcommand{\niiha}{\hbox{\nii/\ha}}
\newcommand{\hbha}{\hbox{\hb/\ha}}
\newcommand{\oiiiha}{\hbox{\oiii/\ha}}
\newcommand{\oiiihb}{\hbox{\oiii/\hb}}

\slugcomment{To appear in {\it The Astrophysical Journal\/}.}

\lefthead{Shopbell \& Bland-Hawthorn} \righthead{Wind in M82}

\begin{document}

\title{The Asymmetric Wind in M82}

\author{P. L. Shopbell\altaffilmark{1}} \altaffiltext{1}{Current
  address: California Institute of Technology, 105-24, Pasadena, CA
  91125} \affil{Rice University, Department of Space Physics \&
  Astronomy, MS 108, Houston, TX 77251; pls@astro.caltech.edu}
\authoraddr{pls@astro.caltech.edu}
\and
\author{J. Bland-Hawthorn}
\affil{Anglo-Australian Observatory, P.O. Box 296, Epping, NSW 2121,
  AUSTRALIA; jbh@aaoepp2.aao.gov.au}

\begin{abstract}
  We have obtained detailed imaging Fabry-Perot observations of the
  nearby galaxy M82 in order to understand the physical association
  between the high-velocity outflow and the starburst nucleus.  The
  high spatial and kinematic resolution of our observations has
  allowed us to perform photometric analyses of \ha, \nii, and \oiii\
  spectral lines at roughly one hundred thousand positions across the
  extent of the galaxy.
  
  The observed velocities of the emitting gas in M82 reveal a bipolar
  outflow of material, originating from the bright starburst regions
  in the galaxy's inner disk, but {\it misaligned\/} with respect to
  the galaxy spin axis. The deprojected outflow velocity indicated by
  the optical filaments increases with radius from 525 to
  655~\kms. All three spectral lines show double components in the
  centers of the outflowing lobes, with the \ha\ line split by
  $\sim$300~\kms\ over a region almost a kiloparsec in size. The
  filamentary lobes lie along an axis tilted by 15\arcdeg\ with
  respect to the spin axis, confirmed by the regions of line splitting
  and the ionization pattern over the outflow. The filaments are not
  simple surfaces of revolution, nor is the emission distributed
  evenly over the surfaces. We model these lobes as a composite of
  cylindrical and conical structures, collimated in the inner
  $\sim$500~pc but expanding at a larger opening angle of
  $\sim$25\arcdeg\ beyond that radius.  We compare our kinematic model
  with simulations of starburst-driven winds in which disk material
  surrounding the source is entrained by the wind. There is some
  evidence for rotation of the wind filaments about the outflow axis
  in support of entrainment, and we find strong similarities between
  the observed and predicted structures.
  
  The data reveal a remarkably low \niiha\ ratio in the region of the
  outflow, indicating that photoionization by the nuclear starburst
  may play a significant role in the excitation of the optical
  filament gas, particularly near the nucleus.  An increase in the
  \oiiiha\ ratio along the outflow is observed. At larger radii, the
  line diagnostics and a strong spatial correlation between \ha\ and
  soft x-ray filaments are consistent with shock ionization.
  
  A smooth spherical halo around M82 is observed in emission lines,
  extending to at least 2~kpc. We propose that the dusty halo is the
  primary source of the linearly polarized optical emission. A diffuse
  ionized medium (DIM) with enhanced \niiha\ emission pervades
  throughout the stellar disk. We discuss likely sources of ionization
  and heating.
\end{abstract}

\keywords{galaxies: individual (M82, NGC 3034); galaxies: starburst;
  ISM: jets and outflows; galaxies: kinematics and dynamics;
  techniques: interferometric}

\section{Introduction}
Ten years ago Chevalier and Clegg (1985) published a brief letter in
which they conjectured that a high supernova rate in a galactic
nucleus could heat the surrounding gas to temperatures with sound
speeds exceeding the escape velocity of the galaxy.  This hot, tenuous
gas would expand outward from the galaxy in the form of a ``wind,''
enriching the surrounding intergalactic medium.  Since that time, the
galactic wind model has been confirmed by observations of over a dozen
galactic-scale outflows: e.g., M82 (\cite{AT78}), NGC~253
(\cite{FT84}), NGC~1569 (\cite{HDLFGW95}), NGC~1705 (\cite{MFD89}),
NGC~1808 (\cite{P93}), NGC~2146 (\cite{AHWL95}), NGC~3079
(\cite{VCBTFS94}), NGC~3628 (\cite{FHK90}), NGC~4051
(\cite{CHSMTGMP97}), NGC~4666 (\cite{DPLHE97}), Mk~509
(\cite{PBAC83}).  These massive ejections of gas and dust are usually
observed as large ($\sim$few kpc), roughly conical structures of
filaments originating in the nuclear regions and oriented along the
minor axes of the galaxies.  They are therefore most often observed in
edge-on disk systems (e.g., M82, NGC~1808), in optical emission lines
(\cite{LH95}), radio continuum (\cite{BODDP93}), and soft X-rays (e.g,
\cite{BST95}; \cite{AHWL95}).  Recent spectral studies have also shown
the ability to detect winds in face-on galaxies using their kinematic
signatures (e.g., NGC~2782 [\cite{BSK92}], Mk~231 [\cite{HK87};
\cite{KCTK97}]).

In some cases these winds are indeed driven by supernovae and massive
stellar winds from a central starburst (e.g., M82, NGC~1569;
\cite{LH95}; \cite{LH96}), while other winds appear to be powered by
more exotic forces associated with the central engines of AGN (e.g.,
NGC~3079, NGC~4051; \cite{CBGOLTMC96}; \cite{CBGOC96}), and some
galaxies appear to exhibit both starburst and AGN characteristics
(e.g., NGC~1808 [\cite{FBW92}], Mk~231 [\cite{LCM94}]).  Because of
the low densities of the wind material, emission from optical lines is
often very difficult to detect, except in nearby galaxies.  The x-ray
emission from these winds is comparable in luminosity to the optical
emission lines, but is visible on larger spatial scales and therefore
detectable to greater distances.

\placetable{m82statstab}

Due to its proximity and favorable inclination angle (see
Tab.~\ref{m82statstab}), the irregular disk galaxy M82 (NGC~3034) has
been studied extensively as a prototype galactic wind system.
Following the original discovery (\cite{LS63}), the first detailed
spectroscopic study of the optical emission line filaments
(\cite{BBR64}) revealed kinematics indicative of a bipolar, roughly
conical, outflow of gas along the minor axis of the galaxy.

Thirty years and over 500 published papers later, the initial
interpretation of the M82 filaments in terms of an outflow still
stands.  Support for this picture extends from radio to gamma ray
wavelengths. The starburst nature of the nucleus of M82 has been
verified through its strong infrared emission (e.g., \cite{SHN87};
\cite{RLSNKLdH88}; \cite{TCJDD91}) and numerous compact radio
supernovae (e.g., \cite{KBS85}; \cite{HTCCY94}; \cite{MPWASd94}).
Models of the starburst evolution produce appropriate quantities of
energy and mass on plausible timescales to create and sustain the
observed nuclear and galactic wind behavior (e.g., \cite{RLRT93};
\cite{DM93}).  The wind itself has been observed optically in both
spectral (e.g., \cite{MHv87}; \cite{MGDP95}) and imaging studies
(e.g., \cite{IvASTY94}), in emission lines (e.g., \cite{AHM90}) and
broadband radiation (e.g., \cite{OM78}).  In particular, kinematic
evidence for the existence of a galactic wind in M82 has been
presented by several optical emission line studies (e.g., \cite{H72};
\cite{AT78}; \cite{BT88}; \cite{HAM90}; \cite{MGDP95}), and even by
molecular observations (\cite{NHHSHS87}).

An extensive x-ray halo has been observed oriented along the minor
axis, extending 5--6~kpc from the disk, far beyond the visible extent
of the optical filaments (e.g., \cite{WSG84}; \cite{SPBKS89};
\cite{BST95}).  A few percent of the supernova energy from the
starburst has been deposited in this hot ($T\sim10^8$~K) ``x-ray
wind.''  A large ($r\sim8$~kpc) spherical halo has been reported at
radio continuum wavelengths (\cite{SO91}), and has been interpreted as
synchrotron emission from relativistic electrons in the outflowing
wind.

With this wealth of data, it is surprising that little attempt has
been made to undertake detailed comparisons of models and
observations.  Large-scale galactic winds were first proposed to
account for the lack of gas in elliptical galaxies (\cite{MB71}).  The
theory of these winds has since evolved in tandem with research on
starburst-driven galactic winds (e.g., \cite{WC83}; \cite{SKC93a};
\cite{SKC93b}).  Since the original starburst wind model
(\cite{CC85}), advances have been made in both analytical studies
(e.g., \cite{KM92a}; \cite{KM92b}) and hydrodynamic simulations (e.g.,
\cite{TI88}; \cite{TB93}; \cite{SBHL94}; \cite{SBHB96}).  Models of
winds in other astrophysical situations, such as stellar winds (e.g.,
\cite{SRLL91}) and winds from AGN (e.g., \cite{S93}; \cite{AL94};
\cite{ALB94}), have contributed to our understanding as well.  On a
broader scale, starburst-driven winds have important ramifications
for a variety of astrophysical and cosmological situations, including
enrichment of the intergalactic medium, contributions to the diffuse
X-ray background, the evolution of dwarf and interacting galaxies, and
the formation of elliptical galaxies through mergers (see \cite{HAM90}
and references therein).

Toward the goal of investigating this model, we have obtained
high-resolution imaging Fabry-Perot observations of M82 in several
optical emission lines.  In Section~\ref{observations} of this paper
we present our Fabry-Perot observations and describe the methods of
data reduction.  In Section~\ref{maps}, we present the derived
two-dimensional emission-line maps of M82 illustrating the
distribution of line flux, ionized gas velocity, and ionization state
across the outflow.  In Section~\ref{discussion}, we describe our new
data for the disk, halo, and outflow and provide comparisons with
other published observations and a number of kinematic models.  We
present our conclusions in Section~\ref{conclusions}.  A subsequent
paper describes refinements to our kinematic and ionization models to
accommodate new observations from the {\it Keck\/} and {\it Hubble
Space Telescopes}.

\section{Observations and Reductions}
\label{observations}

\subsection{Fabry-Perot Observations}
Observations of M82 were obtained in February, 1986 at the 3.6-meter
Canada-France-Hawaii telescope (CFHT) on Mauna Kea, in the emission
lines of \ha\ $\lambda$6563 and \nii\ $\lambda$6583.  The Hawaii
Imaging Fabry-Perot Interferometer (HIFI; \cite{BT89}) was used with
an ET-50 etalon from Queensgate Instruments.  The etalon coatings were
formulated to provide a free spectral range of 85\AA\ and a finesse of
60, for a spectral resolution of 1.4\AA\ (65~\kms, at \ha).  The
$512\times 512$ pixels of the Texas Instruments CCD, after $2\times 2$
on-chip binning, each subtend 0\farcs 86 ($\sim$13~pc at the distance
of M82).  This scale undersampled the estimated 1\arcsec\ seeing disk
but was necessary due to the low flux levels of the outflow emission.
The 3\farcm 5 field covers almost the entire region of minor-axis
optical filaments.  Other characteristics of the observing system are
provided in Table~\ref{HIFItab}, along with their experimentally
determined values.

A second set of Fabry-Perot observations were obtained in March, 1992
at the University of Hawaii 88-inch telescope on Mauna Kea, in the
emission line of \oiii\ $\lambda$5007.  Again the HIFI system was
used, with an ET-70 etalon owned by Dr.\ Sylvain Veilleux.  Although
the free spectral range of this etalon (61\AA) was smaller than that
used for the \ha\ and \nii\ observations, the finesse was essentially
identical, as were the effective spatial and velocity resolutions
(0\farcs 85 pix$^{-1}$ and 60~\kms\ at \oiii, respectively) and the
field of view.  The characteristics of the observing system are
provided in more detail in Table~\ref{HIFItab}.  In both cases, the
etalon was tilted to remove internal reflections and paired with wide
(50\AA\ FWHM) interference filters to eliminate interorder confusion
problems.

\placetable{HIFItab}

\subsection{Data Reductions}
Most of the data reduction was performed under the Zodiac image
processing system (\cite{M82}; \cite{S97}), encompassing standard
routines as well as a number of programs written by the authors and
optimized for Fabry-Perot data.  IRAF\footnote{IRAF is distributed by
the National Optical Astronomy Observatories, which is operated by the
Association of Universities for Research in Astronomy, Inc.  (AURA)
under cooperative agreement with the National Science Foundation.} was
also used for a small number of specific problems.  A summary of the
data reduction steps is now provided; see \cite{S95} for more details.

\subsubsection{CCD Reductions}
The bias level was subtracted from each frame, using the average of
several CCD bias frames.  Cosmic rays were denoted by hand and removed
by nearest-neighbor interpolation.  Bad columns were also denoted by
hand and then removed by sinc interpolation across each row of the bad
column.  The sinc function was selected for use in the interpolation
because of its favorable Fourier properties: its Fourier transform is
the top-hat function, implying even sampling of the image noise
structure over a finite range of frequency.

A flatfield image was constructed for our data by summing a series of
frames of a white source, scanning the etalon through a free spectral
range.  Any first-order slope in the illumination pattern was removed
with a two-dimensional linear fit.  (Sky flats were not available.)
The flatfield was then divided by a smoothed ($\sigma \sim 3$ pixels)
version of itself, normalized, and divided into each data and
calibration frame.  Due to problems with on-chip binning, the movement
of transient defects between the images of each cube, and temperature
fluctuations of the CCD, the flatfielding process did not
substantially alter the noise content of most frames.  Fortunately,
the pixel-to-pixel sensitivity variations comprise a noise source at a
level of only $\sim$1.6\%, a small value compared to other Fabry-Perot
photometry uncertainties.

\subsubsection{Fabry-Perot Reductions}
{\it Cube building.\/} In order to randomize variations in the sky
background due to changing hour angle, approaching dawn, etc., as well
as instrumental fluctuations, the data frames were observed in a
random or staggered order with respect to the etalon optical gap.  The
frames were first sorted by gap and stacked to form cubes: target
(M82) data cubes through each filter, a whitelight or flatfield cube
through each filter, a standard lamp calibration cube, and standard
star cubes.
                
{\it Whitelight calibration.\/} The whitelight cube was used to map
the spatial and spectral shape of the interference filter, which was
then removed in the same manner as flatfielding.  The whitelight cube
was heavily smoothed ($\sigma \sim 15$ pixels) in the spatial
dimension, re-sampled from 0.65\AA\ to 0.99\AA\ steps, to match the
etalon gap spacings of the data cubes, then normalized and divided
into the corresponding data cubes.

{\it Frame alignment.\/} The frames in each data cube were aligned
using two stars near the disk of the galaxy.  (The single bright star
in the field, AGK~3+69~428 [\cite{BG82}], 2\farcm 5 southwest of the
nucleus, saturated the detector.)  Fractional pixel shifts were made
by two-dimensional spline interpolation of the images.  The spatial
registration is accurate to better than one pixel ($\la 0\farcs 5$).

We note here that, although the HIFI system and the tilted etalon
provide a field that is relatively free of ghost reflections, the
location of the bright star AGK~3+69~428 in the southwest corner of
the frame was such that a pair of concentric ghost images of the star
appear opposite the optical axis, in the southeast corner of the
frame.  Following frame registration, a region encompassing the ghost
images was masked from further analysis.  This is unfortunate, in that
the minor axis emission of M82 runs directly through this region, but
it demonstrates the care that must be taken when observing with
Fabry-Perot systems, which always have prevalent ghost patterns.

{\it Ring fitting.\/} Temperature and humidity variations produce
drifts in the the expected spectral response of the Fabry-Perot
etalon.  In order to parametrize the spectral and spatial drifts in
the Fabry-Perot system, we obtained a set of calibration lamp images,
taken periodically throughout the night, all at the same etalon
spacing.  Elliptical fits to the rings revealed no noticeable
variations in the radius or circularity of the rings, indicating that
the spectral stability of the etalon was extremely good.  Flexure of
the telescope system was detectable as a two pixel ($\sim$1\farcs 5)
shift of the ring centers (i.e., the optical axis) over the course of
each night.  The frame alignment procedure has removed this, with
minimal effect along the wavelength axis.

{\it Data cube resampling and smoothing.\/} While our \ha+\nii\ data
was sampled regularly at 0.99\AA\ (45~\kms) intervals across the \ha\
line, observing time constraints required us to interpolate several
frames across the \nii\ portion of the spectra, where the sampling was
a factor of two coarser.  Although a spline interpolation was
performed without difficulty, a slightly larger error should be
assumed for the final \nii\ fluxes and velocities.  The \oiii\ data
set required extrapolation of a single (continuum) frame at either end
of the spectra.

A light Hanning filter was then applied along the spectral axis of
each cube, in order to allow efficient automated fitting of the large
numbers of the spectra.  Tests indicate this had a negligible effect
on the final fit parameters.

{\it Phase calibration.\/} The instrumental profile of a Fabry-Perot
interferometer is a complex function of spatial position, wavelength,
and optical gap spacing, given by the well-known Airy function
(\cite{BT89}).  Due to the large free spectral range and low
interference order of the etalons, the monochromatic ``phase
surfaces,'' as observed in the emission lines of a calibration lamp,
were well parametrized by the analytical expression for the
three-dimensional Airy function.  A fit to this function determined
several system constants listed in Table~\ref{HIFItab}.  This fit was
then used to shift each spectrum in the data cube by the appropriate
value to generate monochromatic frames.  The convergence of two night
sky lines into a single frame each confirmed the accuracy of the phase
correction for the \ha+\nii\ data set.  For the \oiii\ data set,
however, a poorly sampled calibration cube prevented us from
performing an accurate phase correction.  Therefore the \oiii\
observations will be used for flux measurements and morphology, but
not for kinematic studies.

{\it Sky subtraction.\/} A limited field of view prevented us from
obtaining a sky spectrum devoid of galaxy emission.  We therefore
removed the two bright night sky lines in the \ha+\nii\ data set by
subtracting Gaussian components with the proper velocity and a mean
flux level as observed across the field.  The lines were identified as
OH emission at 6553.61\AA\ and 6577.28\AA\ (\cite{OM92}), providing
the wavelength calibration for the spectral axis.  The night sky
continuum was not removed from the data.  The \oiii\ data were not
sky-subtracted, due to the low level of sky flux and the difficulties
associated with the non-phase-corrected data.  Resulting errors in the
final spectral fits are negligible.

\subsubsection{Photometry}
\label{photometry}
Recent studies (e.g., \cite{VCBTFS94}) have shown that Fabry-Perot
data can be flux calibrated to an accuracy equal to that of CCD
imagery and longslit spectroscopy.  Rather than scaling to an
externally calibrated data set, we have applied a primary calibration,
employing standard star observations directly.  Since this Fabry-Perot
flux calibration procedure is described in more detail elsewhere
(e.g., \cite{BSVJ97}), we outline only the salient points below.

The star $\varepsilon$~Orionis was used to calibrate the \ha+\nii\
observations; $\alpha$~Lyrae and $\eta$~Hydrae were used for the
\oiii\ observations.  After cosmetic cleaning and applying the
whitelight correction described above, the counts in each stellar
image were summed, and then scaled by the telescope aperture size and
exposure time.  To obtain pixel values in units of counts cm$^{-2}$
sec$^{-1}$ \AA$^{-1}$ we must then divide by the effective spectral
bandpass of the etalon at each pixel.  We emphasize that this value is
not necessarily the same as either the spectral sampling of the cube
or the etalon resolution.  Rather, it is a measure of the amount of
flux transmitted at a given wavelength for a sequence of etalon
spacings; it depends primarily on the finesse ($N_R$) of the etalon.
This ``effective bandpass'' was calculated by summing the flux under a
synthetic monochromatic spectrum spanning an entire order of the
appropriate theoretical Airy function and dividing by the peak value
of the Airy function.  We derived an effective bandpass of 3.0\AA\ for
our data sets.  The scaled values of observed stellar counts were then
compared with published flux values (\cite{H70}; \cite{HL75}),
corrected for atmospheric extinction.  Each pixel in the final spectra
has units of ergs cm$^{-2}$ sec$^{-1}$ pixel$^{-1}$ frame$^{-1}$,
yielding total fluxes from the line fits in units of ergs cm$^{-2}$
sec$^{-1}$ pixel$^{-1}$.  The final estimated systematic error in the
flux calibration was $\sim$7.5\%.

\section{Fabry-Perot Maps}
\label{maps}
The emission-line spectra were fit with Gaussian profiles across the
field to generate spatial maps of line characteristics such as flux,
velocity, and dispersion.  Single-component fits were made to the \ha\ 
and \nii\ lines across the majority of the field, well into the halo
of the galaxy.  There are regions north of the galaxy's disk where
double components were fit to the \ha\ line profile, as well as
regions south of the disk where double components were fit to both the
\ha\ and \nii\ lines.  The \oiii\ line was fit with a single Gaussian
component throughout the field, although visual inspection revealed
signs of line splitting in specific regions.

The region of acceptable fits is identical for the \ha\ and \nii\ 
maps, although the \nii\ fits in the northern-most regions of the maps
have significantly higher errors, due to clipping of the 6583\AA\ 
profile by the limited spectral coverage.  The region with sufficient
\oiii\ flux for profile fitting is much smaller than that of \ha\ or
\nii.  The central regions of the galaxy are saturated in both the
\ha\ and \nii\ observations.

We now describe the important features of each Fabry-Perot map. Each
spatial map is presented at approximately the same scale, with tick
marks separated by one arcminute ($\sim$950~pc).  The maps from the
\ha+\nii\ data set contain regions in which the spectral lines are
split; these maps show both a flux-weighted total for the galaxy as
well as values for the individual components separately. These
components are referred to as the high-velocity component (HVC) and
the low-velocity component (LVC).  North is up and east is to the left
in all images.  The position angle of the major axis of the galaxy is
approximately 65\arcdeg.

\placefigure{HaNiifluxfig}

Figure~\ref{HaNiifluxfig} illustrates the flux distribution in the
light of \ha\ 6563\AA\ and \nii\ 6583\AA. We measure a total
unsaturated \ha\ flux of $9.9\times 10^{-11}$ ergs cm$^{-2}$
sec$^{-1}$ and a total \nii\ flux of $4.1\times 10^{-11}$ ergs
cm$^{-2}$ sec$^{-1}$.  A rough comparison with the \ha\ imagery to be
presented in the next section indicates that the saturated nuclear
regions contribute an additional $\sim2.1\times 10^{-11}$ ergs
cm$^{-2}$ sec$^{-1}$ in the \ha\ line, or $\sim$17\% of the total.
The flux is concentrated in the nucleus and along the minor axis, with
very little emission originating in the extended disk of the galaxy.
The nuclear line emission has saturated the detector in two
concentrations (``the eyes''), as well as a number of weaker ancillary
regions.  In contrast, the minor axis emission is spatially extensive
and filamentary.  Numerous long radial filaments can be seen extending
more than a kiloparsec from the nucleus, as well as complex
small-scale structures.  Note, for example, the bright bow shock-like
arc approximately 500~pc SSE of the nucleus.

The morphology of the extraplanar gas differs between the two sides of
the galaxy, appearing more chaotic on the north side and showing signs
of collimation in the south.  The southern \ha\ emission exhibits a
sharp edge on the eastern side. Furthermore, the emission to the south
is considerably brighter and more extensive, although there appears to
be an abrupt reduction in flux at a distance of approximately 500~pc
from the nucleus.  Examination of the \ha\ and \nii\ emission maps
reveals that, beyond 500~pc, the flux in the HVC drops rapidly, while
the flux in the LVC remains more uniform.  We also find pervasive
diffuse emission at a low level throughout the halo of the galaxy.

\placefigure{Oiiifluxfig}

Figure~\ref{Oiiifluxfig} is a map of the flux from M82 at \oiii\
5007\AA. We measure a total \oiii\ flux of $2.6\times 10^{-12}$ ergs
cm$^{-2}$ sec$^{-1}$, more than an order of magnitude below that seen
at \ha\ and \nii.  There is essentially no emission from the disk at
this wavelength; the flux originates almost entirely within the
nucleus and along the southern minor axis of the disk.  Although the
radial extent of the minor-axis emission is much smaller than that
seen in \nii\ and \ha, the filamentary morphology is similar.  Indeed,
excellent correlation is seen with the \ha\ flux map, which has been
contoured on Figure~\ref{Oiiifluxfig}.  A striking feature of the
\oiii\ flux map is the presence of two distinct streams of emitting
gas, each extending along the southern minor axis from one of the
bright nuclear emission regions.  These streams can also be identified
in the \ha\ flux map (Figure~\ref{HaNiifluxfig}), although the much
more pervasive nature of the \ha\ emission makes the structure more
difficult to discern at small radii.

\placefigure{Havelfig}

Figure~\ref{Havelfig} is a radial velocity map for the \ha-emitting
gas in M82.  Two prominent trends are seen: first, there exists a
strong velocity gradient along the major axis of the galaxy,
consistent with normal disk rotation, with the eastern end moving away
from the observer.  Second, a velocity transition is evident along the
minor axis of the galaxy: strong blue-shifting of the \ha\ emission is
seen south of the disk; strong red-shifting to the north.  The
major-axis rotation signature merges into this minor axis motion in a
gradual fashion, although this may be due in part to the
flux-weighting procedure.

The HVC consists of highly redshifted emission in the north and highly
blueshifted emission in the south ($v_r\sim300$~\kms).  The LVC
consists of emission at roughly the systemic velocity of the galaxy
($v_{sys}\approx203$~\kms; \cite{ddCBPF91}), in both the north and
south.  The northern and southern component pairs mirror each other
almost exactly in terms of relative velocity and kinematic structure.
Perhaps surprisingly, the velocities within each individual component
exhibit little radial or azimuthal variation.  The exception to this
is the inner portion of the HVC, where the radial velocities approach
the systemic velocity.

A map of the line-of-sight velocities of the \nii-emitting gas was
also produced, but is not presented here, as it provides little
additional information.  The trends along both the major and minor
axes are identical to those seen at \ha, but at a lower
signal-to-noise ratio due to the decreased flux of \nii\ line
emission.  A velocity map of the \oiii-emitting gas was not produced,
for the reasons outlined above.

\placefigure{NiiHalgfig}

Figure~\ref{NiiHalgfig} shows the logarithm of the ratio of the \nii\
6583\AA\ flux to the \ha\ 6563\AA\ flux from M82.  The most remarkable
feature in this map is the presence of a distinct region of low line
ratio (0.0--1.0) along the minor axis, south of the disk.  This region
extends at least a kiloparsec in radius from the nucleus and is
narrower than the regions of minor axis filaments visible in the \ha\
and \nii\ flux maps (Fig.~\ref{HaNiifluxfig}).  The region of low
ratios can be separated into two distinct structures, originating with
the bright central eyes of M82, similar to the structures seen at
smaller radii in the \oiii\ 5007\AA\ flux map
(Fig.~\ref{Oiiifluxfig}).

Elsewhere in the galaxy, higher line ratios ($\gtrsim1.0$) prevail,
particularly at large radii within the inclined disk.  Discrete
\ion{H}{2} regions can be seen in the disk as small concentrations
with the expected \niiha\ ratio of $\sim$0.5 (\cite{O89}).  The
regions of extremely low line ratio to the north are a result of the
incompletely sampled \nii\ line, and should be ignored.

\placefigure{OiiiHalgfig}

Figure~\ref{OiiiHalgfig} is a map of the logarithm of the ratio of the
\oiii\ 5007\AA\ flux to the \ha\ 6563\AA\ flux from M82.  The ratio
has been calculated over the entire spatial extent of the \oiii\
emission, as seen in the flux map (Fig.~\ref{Oiiifluxfig}).  The
observed ratios are low ($\sim$0.05) throughout the nucleus and
regions south of the disk, with the only apparent trend being a
gradual increase in the ratio with distance from the nucleus.

\section{Discussion}
\label{discussion}
We now discuss the observational implications of these data in the
context of each of the galaxy's primary morphological components:
starburst disk, extended halo, and galactic-scale outflow.

\subsection{Starburst Disk}
Our observations reveal a small irregular disk containing heavy
obscuration and a central starburst.  The \ha\ and \oiii\ flux maps
(Figs.~\ref{HaNiifluxfig} and \ref{Oiiifluxfig}) reveal little ionized
gas in the extended disk of M82, outside of the central nuclear
starburst region, as found by slit spectra (\cite{OM78}).  Small
concentrations of line emission that do exist in the disk outside of
the starburst, such as the regions 0\farcm 5 east of the nucleus (see
also \cite{OM78}; Fig.~3), appear to be giant \hii\ regions.  The
\niiha\ ratio map (Fig.~\ref{NiiHalgfig}) indicates moderate values
for these regions ($\sim$0.56), consistent with a cooler range of
photoionized \hii\ regions ($T_{\hbox{max}}\sim40,000$~K, \cite{ED85};
\cite{VO87}).  Such regions exhibit low levels of \oiii\ emission,
consistent with our non-detection in Figure~\ref{Oiiifluxfig}.
Although the high extinction in the disk of M82 (3--27~mag;
\cite{OGHC95}; \cite{MRRK93}) serves to hide smaller \hii\ regions
from our optical observations, the paucity of large \hii\ regions
suggests that the level of star formation must be low outside of the
central starburst.  This conclusion has also been reached by
multi-epoch radio supernovae studies (e.g., \cite{KBS85};
\cite{HTCCY94}).  The weak nature of the \oiii\ emission implies a
high overall metallicity in M82, as has been suggested by previous
observations (e.g., \cite{DEHH87}; \cite{GL92}), and as is anticipated
due to chemical enrichment by the starburst (e.g., \cite{KS96}).

The nuclear emission is dominated by two large saturated regions, each
approximately 200~pc in diameter, and centered $\sim$125~pc from the
2~\micron\ stellar nucleus (see Fig.~\ref{HaNiifluxfig}).  These
regions correspond to knots $A$ and $C$ of \cite{OM78} and are known
to be extremely high surface brightness ``clusters of clusters'' of
young ($T\lesssim50$~Myr) stars (\cite{OGHC95}).  Also identifiable in
Figure~\ref{HaNiifluxfig} are knots $D$ and $E$ (also saturated), as
well as knots $F$, $G$, and $H$.  Knot $B$ is extremely faint at \ha\
wavelengths, especially when compared to broadband (\cite{OM78}) and
ultraviolet (\cite{He96}) observations, suggesting a higher gas
content in the inner regions of the galaxy.  The outflow can be traced
to knots $A$ and $C$.  This relationship is particularly
well-demonstrated by the \niiha\ map (Fig.~\ref{NiiHalgfig}) and the
\oiii\ flux map (Fig.~\ref{Oiiifluxfig}).

\placefigure{rotcurvefig}

We have plotted the \ha\ radial velocities along a straight line
corresponding to the major axis of the galaxy in
Figure~\ref{rotcurvefig} (panel $c$), along with a number of rotation
curves from the literature at a range of wavelengths.  The published
systemic velocity of M82, 203~\kms\ (\cite{ddCBPF91}), corrected to a
heliocentric value of 208.7~\kms, has been subtracted from the \ha\
Fabry-Perot data.  The rotation curve rises to approximately 100~\kms\
within $\sim$9\arcsec\ ($\sim$140~pc) of the nucleus and remains
relatively flat to the edges of the observations.  No substantial
fall-off in the rotation curve is seen, due primarily to the limited
radial extent of line emission from the disk.  The observed rotation
curve matches well those found at \ha\ by \cite{H72} (his Fig.~12) and
\cite{MCGD93} (panel $d$ of Fig.~\ref{rotcurvefig}), including the
turn-over and asymptotic velocities and the nuclear velocity gradient
($\sim$11~\kms\ arcsec$^{-1}$).  Figure~\ref{rotcurvefig} also
illustrates an increase in the central velocity gradient with
wavelength.  As has been pointed out by other authors (e.g.,
\cite{MCGD93}), this effect is indicative of the large extinction
toward the nuclear regions of the galaxy.

The \ha\ rotation curve also correlates well with that of the 250~pc
nuclear ring seen in molecular emission lines (e.g., \cite{LNSWRK90}).
The double-lobed structure of the central \ha\ emission is suggestive
of an edge-on ring structure as well, interior to the molecular ring.
The starburst could be identified with the ring of ionized gas and is
probably propagating outward, fueled by the cold gas in the molecular
ring (\cite{WGT92}; cf.\ \cite{SL95}).  This ring may also provide
much of the extinction toward the nucleus of M82 (e.g., \cite{TG92}).
The dynamic conditions of the central regions would have removed most
of the obscuring material interior to the ring, as has been suggested
in our Galaxy (e.g., \cite{BGW82}).  The resolution of the central
star clusters by recent HST observations (\cite{OGHC95}) suggests that
any such ring would be very clumpy however, and the two bright regions
may indeed represent real spatial enhancements in the distribution of
ionized gas and star formation.  The relationship between this ring
structure and the proposed bar in M82 ($r\sim500$~pc; \cite{TCJDD91};
\cite{AL95}) is not clear.

The \niiha\ line ratio map (Fig.~\ref{NiiHalgfig}) shows a high ratio
in the disk, especially at large radii, where the value is observed to
exceed 1.0.  The actual line ratios in the disk may be even slightly
higher and possibly more uniform, due to dilution by the central
starburst and the halo.  Such high ratios are by no means rare in the
nuclei of disk galaxies (\cite{K83}), particularly LINERS (e.g.,
NGC~4319 [\cite{SA87}], NGC~5194 [\cite{FCJLV85}; \cite{GG85}]), and
are understood to originate with shock excitation and/or
photoionization by a power-law source (\cite{VO87}).  Although it is
unlikely that M82 harbors an AGN (e.g., \cite{RLTLT80}; \cite{CP92};
\cite{MPWASd94}), a comparison with other galaxies which exhibit
pervasive high \niiha\ line ratios in their extended disks is
instructive: studies of NGC~3079 (\cite{VCBTFS94}) and especially
NGC~1068 (\cite{BSC91}) have found high \niiha\ ratios of 0.6--1.3
across much of the inner disk.  In NGC~1068, the disk \hii\ regions
are found to reside in regions of lower \niiha\ ratio
($\sim$0.3--0.8), while the disk as a whole is permeated with gas
exhibiting the higher ratios, similar to what we observe in M82.  In
order to boost this forbidden line to recombination line ratio, the
``heating per ionization'' must be high, for which most models require
high energy photons, energetic electrons, or a dilute radiation bath
(\cite{S92}).

In the case of M82, an attractive candidate to enhance the \niiha\ 
ratio in the disk is the concept of ``mixing layers'' (\cite{SC93};
\cite{VDS94}). The turbulence resulting from the interaction of hot
supernovae remnants with the ambient ISM creates an intermediate
temperature phase, which models suggest emits a radiation field
somewhat harder than thermal bremsstrahlung.  This process can produce
\niiha\ ratios in excess of 1.0 (\cite{SSB93}).  By employing soft
X-rays as the photoionization mechanism, mixing ratios have been used
to model line emission ratios of up to 3--4 in cooling flows
(\cite{DV91}; \cite{CF92}).  Considering the current optical
appearance of the disk of M82, its interaction with the galaxy M81
approximately $10^8$ years ago, and the energetic activity associated
with the nuclear starburst, a turbulent inner disk would not be
unexpected.

However, as mentioned above, our observations do not detect
significant numbers of star forming regions outside the starburst
nucleus.  Even if this is attributed to high levels of extinction in
the disk, other wavebands confirm the low level of star formation. All
of the radio supernovae discovered by \cite{KBS85} are within 300~pc
of the galaxy's nucleus, well inside the region of highest \niiha\ 
ratios.  The diffuse X-ray flux in the disk has also been shown to
decrease rapidly with radius, implying a reduced star formation rate
outside the nuclear regions (\cite{BST95}).  Other models for
producing high \niiha\ ratios in the disk must be considered, such as
chemical enrichment and cosmic ray heating.  A combination of shock
and photoionization may also provide a solution (e.g., \cite{HG90}),
although detailed models are not yet available.

\subsection{Extended Halo}
Early polarization observations detected a strong linear polarization
throughout the halo of M82, with the position angles oriented
perpendicular to a radial vector from the nucleus (\cite{E69}).  This
polarization was interpreted as evidence for a scattering component in
the galaxy (e.g., \cite{S69}), probably consisting of electrons
illuminated by the nucleus.  The more difficult issue has been the
polarization of the optical filaments themselves.  The first
observations in this regard (\cite{VS72}) determined that the minor
axis filaments and the halo were equally polarized at optical
wavelengths, suggesting that perhaps there was no ``explosion'' in M82
and that the off-axis filaments were merely density enhancements in a
dusty cloud through which the galaxy was moving (\cite{SMM77}).

Although the discovery of split emission lines (\cite{AT78}) and the
minor-axis X-ray halo (\cite{WSG84}) have virtually eliminated this
alternate interpretation for the optical filaments (although see
\cite{RMS87}), the polarization measurements remain poorly understood.
Recent observations (e.g., \cite{SEA91}) indicate that the optical
filaments may indeed contain a scattered component, but even then
there is uncertainty as to the source of the polarization, i.e. the
nucleus (\cite{V74}) or the entire disk (\cite{SM75}).  Unfortunately,
with few exceptions (e.g., \cite{SAC76}; \cite{BT88}; \cite{D93};
\cite{SBHL94}), the presence of a halo component in M82 has been
largely ignored.

Our detailed analyses in \ha\ and \nii\ clearly confirm the existence
of the smooth exponential halo noted by \cite{BT88}.
Figure~\ref{exphalofig} illustrates the flux along a narrow
($\sim$9\arcsec) band parallel to and approximately 45\arcsec\
southeast of the major axis of the galaxy.  The flux due to the
outflow filaments can be seen superimposed upon an exponentially
decreasing background.  The halo has be detected across our entire
region of fit lines, approaching a flux level of $10^{-15}$ ergs
cm$^{-2}$ sec$^{-1}$ arcsec$^{-2}$ at a radius of 1~kpc.

\placefigure{exphalofig}

We have estimated the radial profile of the halo flux at \ha\ with an
exponential function:
\begin{equation}
        I(\hbox{\ha}) = I_0(\hbox{\ha}) e^{-r/r_e},
\end{equation}
where $I_0(\hbox{\ha}) \sim 1.6\times10^{-15}$ ergs cm$^{-2}$
sec$^{-1}$ arcsec$^{-2}$ \AA$^{-1}$ and $r_e \sim 315$~pc.  This
profile has been overlayed on the cut in Figure~\ref{exphalofig}.  For
the observed halo line width of $\sim$350~\kms, the integrated \ha\
flux is $\sim 2.5\times10^{-11}$ ergs cm$^{-2}$ sec$^{-1}$.  This
compares with an observed flux from the filaments and {\it
unsaturated\/} nucleus of $\sim 9.9\times10^{-11}$ ergs cm$^{-2}$
sec$^{-1}$.  For a distance of 3.25~Mpc this implies a total \ha\
luminosity from the halo of $3.2\times10^{40}$ ergs sec$^{-1}$.

The azimuthally symmetric polarization pattern of the halo in
broadband light (\cite{SAC76}) suggests a scattering origin.  The halo
is known to comprise cold neutral atoms (\cite{C77}), relativistic
electrons (\cite{SO91}), dust (\cite{VS72}) and warm ions
(\cite{BT88}).  If we associate the line-emitting halo with the
polarized component, the line width ($\sim$350~\kms\ FWHM) reflects
either the motion of scattering mirrors embedded in a warm medium or
the ``beam-averaged'' projected kinematics of the nuclear and
large-scale disk gas. If the observed line dispersion arises from
thermal motions of electrons, the kinetic temperature must be less
than 1000 K, at which point the flux from recombination would
overwhelm the scattered flux for any reasonable halo density ($n_H
\approx 1$ cm$^{-3}$; \cite{C77}). Dust scattering is expected to be
much more efficient in any case. The ratio of the scattering optical
depths can be written
\begin{equation}
        R_{de} = {{\sigma_d n_d}\over{\sigma_e n_e}},
\end{equation}
where $n_d$ and $n_e$ are the dust and electron densities; $\sigma_d$
and $\sigma_e$ are the dust and electron scattering cross sections.
For $\sigma_d$, we form the more conservative weighted mean of the MRN
(\cite{MRN77}) grain distribution for which $\sigma_d \approx
0.01\mu$m. In the local ISM, the dust-to-gas {\it number\/} density
ratio is $n_d / n_H \approx 10^{-12}$ (\cite{OS73}) although this
depends strictly on the grain composition. With the conservative
assumption of a totally ionized halo, we deduce a scattering ratio
$R_{de} \sim 50$, verifying the relative importance of dust
scattering.  \cite{SAC76} demonstrate that a reasonable optical depth
to dust scattering is $\tau_d \sim 0.1$ from a comparison of the halo
broadband flux and the estimated disk flux.

The halo dust could reside in an extended neutral or warm ionized
medium, some fraction of which could be supplied by the energetic wind
(\cite{BBR64}).  We now compare the timescale for dust destruction by
sputtering with the estimated age of the starburst wind, $\tau_{wind}
\sim 3 \times 10^6$ years (e.g., \cite{LS63}; \cite{BT88}).  From
\cite{OS73}, the timescale for grain sputtering is
\begin{equation}
    \tau_{sput} \sim 3 \times 10^5\ n_e \left({{10^6}\over{T_e}}\right)^{1/2}
    \left({{0.01}\over{Y}}\right) \left({{r_d}\over{0.1}}\right) \quad
    \hbox{years,}
    \label{sputeq}
\end{equation}
where $Y$ is the sputtering ``yield,'' i.e., the number of atoms
released per impact.  $Y$ varies from $\sim 5 \times 10^{-4}$ at the
threshold sputtering temperature of $3 \times 10^5$ K to $\sim 0.01$
for temperatures of $10^7$--$10^8$ K.

At the low temperatures expected in a galactic halo,
$T\lesssim10^5$~K, the lifetime of dust is greater than $10^7$ yrs,
comparable to or longer than the lifetime of the outflow.  However,
any dust located directly in the hot ($T\sim10^8$~K) X-ray-emitting
wind survives for no more than $10^5$ years, and has therefore been
destroyed.  Such an effect is supported by low-resolution radio maps
of the M81/M82 region, which indicate an anti-correlation between the
wind lobes and \hi\ column density (\cite{C77}).  We note that more
recent studies of gas-grain sputtering and grain-grain collisions
(\cite{TMSH94}; \cite{JTH96}) suggest that grains may be able to
survive much longer than previously thought.  However, most of these
studies are specific to the three-phase ISM in the Galaxy, and it
remains unclear precisely how the results should be extended to
galactic wind systems, which characteristically involve higher
temperatures ($10^8$~K vs.\ $10^6$~K) and larger velocities (600~\kms\
vs.\ 200~\kms) than standard ISM models.

Given the sputtering timescales, if the dust has been delivered into
the halo by the wind, it must be as a component of cooler material
entrained by the hot wind itself.  However, this implies that the halo
dust and optical emission line filaments may then have the same origin
and morphology, yet the polarization observations do not seem to
indicate a minor-axis concentration in the dust distribution
(\cite{SM75}; \cite{SAC76}; \cite{SEA91}).  We also do not observe
substantial redshifted emission south of the galaxy, as would be
expected from mirrors moving with the outflow.

Alternatively, dust may have been forced into the halo at an early
stage of the outflow, at a time when it was dominated much more by
radiation from massive stars in the central burst than by supernovae.
Such a mechanism has been hypothesized by \cite{F97} to explain the
appearance of high-z dust in the Galaxy and other edge-on spirals.  It
has been demonstrated that radiation pressure is sufficient to
evacuate a large fraction of the dust near an active star-forming
region into the halo, creating a dust distribution which varies slowly
with height above the disk.  Given the current importance of radiation
effects in the inner portion of the M82 outflow (see
Section~\ref{compoptxray} below) and the extensive nature of the
observed dusty halo, such a scenario seems a reasonable model.

Regardless of any mechanism of relocating disk dust into the halo,
clearly the hypothesized encounter between M81 and M82
$\sim10^8$~years ago (\cite{C77}; \cite{YHL94}) has played an
important role in the evolution of the halo in M82.  (This encounter
is also thought to have initiated the central starburst in M82, e.g.,
\cite{MH94}.)  Radio observations have shown massive clouds of \hi\
surrounding both galaxies, with large arcs and bridges joining them
and the nearby galaxy NGC~3077 (e.g., \cite{D74}; \cite{YHL94}).  The
large-scale velocity structure of this gas blends with the global \ha\
velocity trends in M82, matching the systemic velocity and even the
minor-axis velocity gradient (\cite{C77}).  The \hi\ cloud is clearly
extensive enough to replenish dust that has been destroyed by
sputtering.  Moreover, this massive reservoir of atomic gas should
help to maintain the tenuous halo gas itself in M82, a component that
is required by hydrodynamical models in order for the outflowing wind
to produce observable structures (\cite{SBHL94}).

\subsection{Galactic-scale Outflow}
A number of studies have concentrated on the minor axis filaments in
M82. Imaging observations (e.g., \cite{LS63}; \cite{IvASTY94}) have
been directed toward understanding the morphology of the outflow,
while spectral studies (e.g., \cite{BBR64}; \cite{H72}; \cite{AT78};
\cite{WCS84}; \cite{BT88}; \cite{MGDP95}) have attempted to
parametrize the kinematics of the filaments.  The most recent
observations detect split emission lines that suggest a pair of
expanding bubbles or cones.  The exact size and shape of the expanding
structures remains a topic of some debate.  Additional optical
spectroscopy has measured the radial variations of line emission along
the outflow (e.g., \cite{HAM90}).  Ratios such as \niiha\ increase
with distance from the nuclear starburst, suggesting a gradual change
in the gas excitation mechanism.  We now investigate these findings in
light of the spatial and spectral coverage provided by our Fabry-Perot
spectrophotometry.

\subsubsection{Optical Morphology}
Optical emission line images of the M82 outflow reveal a complex
distribution of radial filaments and knots along the minor axis of the
galaxy.  In order to examine the morphology of these filaments, R. B.
Tully kindly obtained for us very deep ($t\sim3000$~sec) \ha\ imagery
at the University of Hawaii 88-inch telescope on Mauna Kea
(Fig.~\ref{deephafig}).  We imaged a 6\arcmin\ circular field, a
factor of two larger than our Fabry-Perot field.  The optical
filaments associated with the minor axis outflow are visible across
the entire field, to radial distances of at least 3.5~kpc to the north
and 2~kpc to the south.  This represents approximately half the extent
of the soft X-ray halo (see below).  The filaments are bright within a
kiloparsec of the disk, particularly in the south.  In these inner
regions, a web of long ($\sim$300~pc) filaments appear distributed
roughly parallel to the minor axis, amidst a ``foam'' of smaller
emission structures.  Beyond a kiloparsec in radius, the filamentary
structure breaks up into fainter, more isolated clumps.

\placefigure{deephafig}

The faintness of the northern filaments is almost certainly due to
obscuration by the inclined disk of the galaxy (cf.\ \cite{He96}).
The inclination angle of M82 is estimated to be 81\fdg 5
(\cite{LS63}), such that the nearer edge of the galaxy is projected on
the northwest side of the nucleus.  We observe the nuclear regions of
the galaxy through the southern side of the disk, and therefore expect
the inner filaments to be much brighter there.  If we assume a
line-of-sight dimension for the optical disk equal to its linear
dimension on the sky ($\sim$11\farcm 2; \cite{ddCBPF91}), we expect
the northern outflow to be at least partially obscured to projected
radii of $\sim0\farcm 8$.  This is consistent with the similar
morphologies of the filaments beyond this radius in the north and
south in Figure~\ref{deephafig}.

Although complicated by obscuration in the north, the inner filament
structure also appears to differ between the two lobes in terms of
collimation.  The bright inner kiloparsec of the southern lobe is
relatively well confined to the minor axis, whereas the northern
outflow filaments cover a much wider range of opening angle.  This is
seen particularly well in the \ha\ flux map from the Fabry-Perot data
(Fig.~\ref{HaNiifluxfig}), and is probably not an obscuration effect,
but rather a difference in the physical morphology of these two lobes.
One possible interpretation is that the nuclear starburst is located
slightly above the galactic mid-plane.  The smaller mass of covering
material to the north would make collimation of an expanding wind more
difficult, resulting in an almost immediate ``breakout'' of the wind
from the disk.

Finally, we have noted that the \niiha\ ratio map and \oiii\ flux maps
indicate that the southern outflow involves two distinct components,
each originating from one of the central bright emission regions.
This may be a ``limb brightening'' effect as has been suggested for
the pair of central emission regions themselves.  This implies that
the emitting filaments are distributed along the outer surface of the
outflow, rather than throughout the volume.  On the other hand, it is
likely that the two bright central regions from which the outflow
streams appear to originate are stellar ``superclusters''
(\cite{OGHC95}) which merely happen to be physically located on either
side of the kinematic center of the galaxy, from our point of view.
Regardless, we do not observe emission enhancement along the outflow
axis, as would be the case for a volume brightened distribution.

If we do not include emission arising within approximately 8\arcsec\
(125~pc) of the disk, our line fits encompass an \ha\ flux of
$7.6\times10^{-11}$~ergs s$^{-1}$ cm$^{-2}$.  After a rough
subtraction of the halo model given in the previous section, this
implies a total filament luminosity
$L(H\alpha)\sim7.6\times10^{40}$~ergs s$^{-1}$ (cf.\
Tab.~\ref{m82statstab}).  Assuming the filaments are completely
ionized, using a case B recombination coefficient for $T\sim 10,000$~K
of $\alpha_B = 2.59\times10^{-13}$~cm$^3$ s$^{-1}$ (\cite{O89}), and
employing the outflow geometry to be discussed in
Section~\ref{kinmodsec} below, we calculate an rms filament density
$\left<n_e\right>\sim5\cdot f^{-1/2}$~cm$^{-3}$.  A very rough
estimate for the filament filling factor of $f\sim0.1$ would suggest a
mean electron density in the filaments
$\left<n_e\right>\sim15$~cm$^{-3}$, easily consistent with \sii\
doublet ratios in the low-density limit throughout much of the large
volume of outer filaments.  This mean density implies a filament mass
$M\sim5.8\times10^6$~M$_{\sun}$ distributed in a volume
$V\sim1.1\times10^8$~pc$^3$, and a kinetic energy in the ionized
filaments $KE\sim2.1\times10^{55}$~ergs.  These values are consistent
with those first computed by \cite{LS63}, $5.8\times10^6$~M$_{\sun}$
and $2.4\times10^{55}$~ergs, respectively.  The entire mass of
outflowing gas is estimated to be a couple orders of magnitude larger
(\cite{HAM90}).  Note that the kinetic energy in the filaments is only
about one percent of the estimated input supernovae energy ($\sim 2
\times 10^{57}$~ergs; \cite{WSG84}).

\subsubsection{X-Ray Morphology}
A large x-ray halo has been observed in M82 (\cite{WSG84};
\cite{SPBKS89}; \cite{TOMMK90}; \cite{BST95}), extending at least 5--6
kiloparsecs in radius along the minor axis of the galaxy.  Originally
believed to be thermal emission from the hot ($T\sim10^7$--$10^8$~K)
gas within the outflow, current models suggest an origin in
wind-shocked clouds. J.N. Bregman has kindly provided us with a
recently published (\cite{BST95}) image of M82 taken with the {\it
R\"ontgensatellit\/} ({\it ROSAT}; \cite{T84}) High-Resolution Imager
(HRI; \cite{PBHKMPRSZC87}).  The pair of 13\arcmin\ square images with
total exposure time 33.8 kiloseconds were taken in early 1991 and late
1992.  The resolution is about 5\arcsec.

Contours representing the {\it ROSAT\/} soft x-ray flux have been
overlayed on our deep \ha\ image (Fig.~\ref{deephafig}).  The
morphologies are markedly similar, with specific filaments and knots
clearly enhanced in both emission bands.  (Note, for example, the two
extensive filaments in the north, and the knots approximately 500~pc
SW of the nucleus.)  But in light of the more extensive distribution
of the X-rays versus the optical emission, it seems unlikely that {\it
all\/} of the X-ray flux is associated with specific filaments of
optical emission. Since the regions of high optical/X-ray correlation
are also some of the brightest in \ha, these knots presumably
represent density enhancements in the outflow bubbles. The large
extent of the X-ray halo and close spatial correlation with the \ha\
emission suggest an interpretation in terms of shocks driven by a
fast, rarefied wind plowing into denser halo gas (e.g.  \cite{SBD93};
\cite{DS96}).  Our observed \ha/X-ray luminosity ratio of $\approx$30
in the brightest filaments, identical to that derived by \cite{PC96},
suggests a similar scenario.  We defer detailed modeling of the \ha\
and soft x-ray emission to a subsequent paper, in which the shocks
will be constrained by deep spectroscopy and imaging from the {\it
Keck\/} and {\it Hubble Space Telescopes}.

\subsubsection{Optical Kinematics}
Our Fabry-Perot observations provide velocity information across the
entire outflow in M82.  Two well-known features of this emission line
gas should be emphasized {\it a priori}.  First, the velocity
signature of outflow is clearly observed along the minor axis in both
\ha\ and \nii\ (see Fig.~\ref{Havelfig}).  The well-established disk
inclination angle of 81\fdg 5 (\cite{LS63}) and the increased opacity
toward the northern filaments clearly establishes the large-scale
kinematics as arising from outflow (cf.\ \cite{E72}; \cite{SMM77}).
Second, the optical emission lines split into two distinct velocity
components in large regions along the minor axis of the galaxy.  The
velocity structure of the individual components, as well as the
difference between them, provides convincing evidence for the presence
of outflowing gas along the minor axis of M82 (\cite{AT78}).
Figure~\ref{Hamosaicfig} illustrates the velocity structure of the
\ha-emitting gas in M82, as observed in the frames of the Fabry-Perot
data cube.  Both minor-axis outflow lobes span velocity ranges
exceeding $v_{proj}\sim300$~\kms, overlapping at the systemic
velocity.

\placefigure{Hamosaicfig}

We now describe our attempts to understand the complex dynamics of the 
outflow in M82 with the aid of kinematic models.  We have extracted 
one-dimensional velocity cuts and synthetic two-dimensional spectra 
from the Fabry-Perot data cubes.  These cuts have been made relative 
to the outflow axis, at an estimated position angle of 
$\sim$150\arcdeg\ ($\sim$85\arcdeg\ relative to the major axis of the 
galaxy; \cite{MGDP95}).  While some authors have argued for a 
spherical wind in M82 (e.g., \cite{SO91}), most recent models find 
evidence for a bipolar outflow, usually in the shape of cones with 
opening angles much less than 90\arcdeg.  This is borne out by our 
study: the line flux, splitting, and ratio maps all exhibit marked 
azimuthal variations indicative of an aspherical outflow morphology, 
strongly weighted toward the minor axis of the galaxy.  Studies of the 
supernova distribution (e.g., \cite{KBS85}) and other disk 
observations (e.g., \cite{AL95}) indicate that even the central 
injection zone is probably not spherical, but rather in the form of a 
flattened disk, with dimensions $\sim$600~pc wide and $\sim$200~pc 
thick.

The specific velocities predicted by \cite{CC85} do not agree with our
observations.  An estimated supernova rate of 0.3~yr$^{-1}$
(\cite{RLTLT80}) yields an outflow velocity from their model of
2000--3000~\kms\ at the edge of the starburst injection zone.  This
value is several times larger than the deprojected velocities observed
at optical wavelengths ($v\sim$525--655~\kms; see below) on much
larger scales.  However, the extensive radio continuum halo
(\cite{SBB85}; \cite{SO91}) could arise from synchrotron radiation due
to a population of relativistic electrons, as they are transported
outward in a wind at velocities in the range considered by Chevalier
\& Clegg.  The relationship between such a wind and the slower, denser
minor-axis outflow seen at optical and x-ray wavelengths remains
unclear. We suspect that much of the \ha\ filamentation arises from
large-scale shocks from a high-speed wind plowing into the gaseous
halo and entrained disk gas. Only a small fraction of the total wind
energy is encompassed by the radio halo ($\sim$2\%; \cite{SO91}).

\placefigure{windvelfig}

In modeling the bipolar outflow in M82, we model the individual line
components rather than flux-weighted velocity profiles.  In
Figure~\ref{windvelfig}, we show the velocities of line fits to the
dual \ha\ components along the minor axis of the galaxy.  While the
northern outflow components show a relatively constant projected
separation of $\sim$300~\kms, the southern region of split lines
reveals an intriguing variation.  Within 200~pc south of the nucleus,
where the favorable inclination of the disk allows us to measure line
profiles closer to the starburst, the individual components of \ha\
are separated by a much smaller velocity, comparable to our resolution
($\sim$50~\kms).  Between 200 and 500~pc from the nucleus, the
components rapidly diverge, remaining at a constant separation of
$\sim$300~\kms\ beyond 500~pc.  Maps of the line component splitting
reveal a separation of this order throughout the spatial extent of
both lobes.  Also plotted in Figure~\ref{windvelfig} are fits to the
\ha\ profiles from the long-slit optical spectra of \cite{MGDP95},
which clearly confirms our observed minor-axis trend in the line
splitting.  An identical radial trend is seen in the \nii\ line
components along the southern outflow axis, but is not shown in the
figure.

\subsubsection{Kinematic Models}
\label{kinmodsec}
In order to understand the intrinsic velocity structure underlying the
observed kinematics, we have undertaken two separate sets of
three-dimensional Monte-Carlo wind simulations: a rounded expanding
bubble and a pair of cones arranged as a funnel.  Although primarily
geometric in nature, these models allow us to estimate the
three-dimensional morphology of the outflow, as well as intrinsic gas
velocities, both of which are important constraints for physical
models of the wind emission mechanisms and other observational
phenomena.

\placefigure{modelsfig}

For our initial model, we used rounded bubble geometries, given by the
spherical functions,
\begin{eqnarray}
  \rho &=& A\ \cos{m\theta} \qquad -\frac{\pi}{2m} < \theta < +\frac{\pi}{2m}\\
  \rho &=& A\ \cos^m \theta \qquad -\frac{\pi}{2} < \theta < +\frac{\pi}{2}
  \label{bubblesimeq}
\end{eqnarray}
(see Fig.~\ref{modelsfig}$a$). Both functions produce a parabolic
leading surface becoming conical at small radii. The parameter $m$ can
be related to an opening angle and a Mach number in cases where the
leading surface is produced in a bow shock.  Truncating the bubble at
inner and outer radii allows for the selection of a outflow structure
with specific opening angle and radial curvature.  We superimposed two
alternate velocity laws upon this spatial geometry: a radial velocity
vector, which corresponds physically to the case in which each gas
parcel is accelerated by the flow originating at the bubble apex, or a
velocity vector perpendicular to the surface of the bubble, which may
be more appropriate for an expansion due to increasing heat and
pressure inside the bubble, such as an inflating balloon.

We then executed a series of Monte-Carlo simulations for each velocity
law, varying bubble parameters such as opening angle, inclination
angle, and inner and outer truncation radii.  The conclusion reached
from this set of simulations is that {\it the rapid divergence of the
velocity components in the southern outflow cannot be reproduced by a
single bubble}, at least not without invoking highly contrived
velocity profiles for the wind.  Studies of the minor axis x-ray
distribution are similarly unable to model the emission with a single
bubble or cone (e.g., \cite{SBHB96}).

Morphologically, one can divide the southern region of split lines
into two separate velocity regimes: the region within approximately
200~pc of the nucleus, where the split line components are separated
by $\sim$50~\kms, and the region beyond 500~pc radius, in which the
components are separated by a much larger, but still relatively
constant, projected value of $\sim$300~\kms.  The inner component is
not observed north of the galaxy, presumably because the split lines
cannot be resolved at a sufficiently small radius due to the
intervening inclined disk of the galaxy.  Note that the \ha\ and \nii\
flux maps (Fig.~\ref{HaNiifluxfig}) also suggest the presence of two
distinct regions, as the line flux drops sharply at the same radius at
which the velocity components rapidly separate ($\sim$500~pc).

We therefore performed another set of Monte-Carlo simulations, this
time using a double cone geometry, given by the cylindrical function
\begin{equation}
  r = \cases{\kappa_1 (z - z_{01}), & 0 $<$ z $<$ 350~pc \cr
             \kappa_2 (z - z_{02}), & 350 $<$ z $<$ 800~pc\cr}
  \label{conesimeq}
\end{equation}
(see Fig.~\ref{modelsfig}$b$).  The $\kappa$ parameters determine the
opening angles of the cones, while the $z_0$ parameters control the
radial extent of the cones through truncation.  Again, we superimposed
both radial and normal (i.e., perpendicular) velocity laws, but
determined in the end that a velocity vector tangential to the cone
surfaces provided the best match to the observations and was most
easily understood physically, in terms of material entrained by the
high-velocity wind.  An additional parameter was used to smooth the
abrupt projected velocity transition where the two cones meet (at
$z=350$~pc).

\placefigure{conesfig}

\placetable{conetab}

% Cone opening angles:
After performing simulations over a range of cone sizes, inclination
angles, opening angles, and velocity laws, we derived the final model
shown in Figure~\ref{conesfig}.  Parameters of the cones which best
fit the observations are given in Table~\ref{conetab}. The inner
``cone'' is almost a cylinder of radius equal to the injection zone
($r\sim200$~pc), with a small, but non-zero, opening angle.  The outer
cone has an opening angle of approximately 25\arcdeg, in relative
agreement with the ``small'' opening angle models of \cite{BT88} and
\cite{MGDP95}.  Models in the ``large'' opening angle regime (e.g.,
60\arcdeg; \cite{HAM90}) do not match the observations, requiring
excessively low intrinsic velocities and a larger projected spatial
extent for the outflow region (see Fig.~\ref{conesfig}$a$).  A large
opening angle for the primary outflow cone also produces substantially
skewed Doppler ellipses in spectra perpendicular to the outflow axis.
This effect is due to the slit cutting through the back and front of
the cone at different nuclear radii, and is not observed in our
synthetic spectra (see Figs.~\ref{conesfig}$c$ and \ref{conesfig}$d$).

% Cone inclination angles:
The inner and outer cones are inclined toward the observer by
$\sim$5\arcdeg\ and $\sim$15\arcdeg, respectively, roughly aligning
the back sides of the two cones (see Fig.~\ref{modelsfig}$b$).  This
is required to explain the lack of a sharp velocity gradient in the
low-velocity component (LVC) at a radius of 350~pc, as is observed in
the high-velocity component (HVC).  In addition, since the LVC
exhibits small but non-zero projected velocities, the back sides of
the cones must be at a slight angle to the plane of the sky.  These
inclination angles agree with previous estimates (e.g., \cite{BBR64};
\cite{He96}).  While the observed velocities of the HVC could be
duplicated with smaller cone opening angles and a larger inclination,
the low velocities of the LVC require small inclination angles.

% Cone velocity laws:
Initial attempts to model the kinematics of the outflow with a
constant velocity law, i.e., using only the simple double-cone
geometry to reproduce the observed velocity structure, were not
successful.  Figure~\ref{windvelfig} illustrates that both velocity
components in the south and north exhibit non-zero slopes in the
position-velocity plot.  This can be understood as either a continuous
change in the intrinsic gas velocity or as a change in the projected
velocity through a continuous change in the outflow cone geometry.
The latter case implies that both sides of the outer cone are
constantly bending toward the observer, producing a slowly increasing
projected velocity with radius. A gradually increasing intrinsic wind
velocity is probably necessary as well, and can be understood from a
physical standpoint.  Buoyancy effects in the hot wind from the
decreasing disk density with scale height, decreasing wind densities
from the lack of collimation at larger radii, and other effects
contribute to produce wind velocities that increase with radius in
standard galactic wind models (e.g., \cite{CC85}; \cite{SBHL94}).
After testing a number of stronger power-law expressions for the gas
velocity dependence on radius, we finally chose the simple linear
model given in Table~\ref{conetab}.  Together with a constant cone
opening angle, this intrinsic velocity structure produces linear
projected velocity gradients that correspond well to those observed in
Figure~\ref{windvelfig}.  The intrinsic velocities of the gas range
from 525~\kms\ near the nucleus to 655~\kms\ at a radius of 1~kpc.
These velocities are comparable to the escape velocity for M82 (see
Table~\ref{m82statstab}), implying that the most distant entrained
filaments are not bound to the galaxy.  This conclusion is also
supported by the large radial extent of the fast wind itself, as seen
in soft X-rays, along the minor axis.

% Cone position angles:
Just as the outflow cones are not aligned with the minor axis of the
galaxy along the line of sight, neither are they aligned in the plane
of the sky.  The region of split \ha\ lines constitutes a cone on the
sky with the expected opening angle of $\sim$25\arcdeg, but with a
position angle of $\sim$165\arcdeg, approximately 15\arcdeg\ greater
than what the literature (e.g., \cite{MGDP95}) had previously defined
as the ``outflow axis.''  The cone axis is rotated $\sim$100\arcdeg\
from the major axis of the galaxy, and places the eastern edge of the
cone almost directly parallel to the minor axis.  In fact,
Figure~\ref{conesfig}$a$ illustrates that this eastern edge is quite
pronounced, both in the split emission lines and in the \ha\ flux
observed from the inner collimated zone.  This suggests that the
tilting of the outflow cones in the plane of the sky has been produced
by a relative density enhancement in the eastern lower halo which
maintains collimation of the wind even as it fans out toward the west
and toward the observer.

While the large outflow cones appear to originate east of the galaxy's
minor axis, a small region of split lines is also observed on the
western edge of the collimated zone, approximately 300~pc from the
nucleus.  The morphology of the \ha\ and \oiii\ flux maps
(Figs.~\ref{HaNiifluxfig} and \ref{Oiiifluxfig}) indicates that this
region constitutes a small bubble on the side of the larger outflow
structure.  We clearly see an enhanced rim of \ha\ emission around the
bubble, and split lines within it.

Recalling again the identification of two outflow ``streams'' in the
\oiii\ flux map and the \niiha\ ratio map (Figs.~\ref{Oiiifluxfig} and
\ref{NiiHalgfig}), one might expect a more substantial outflow from
the western half of the nucleus, or at least a more
centrally-positioned outflow cone.  However, it appears that gas
densities on the western side of the lower halo are substantial enough
to keep the western stream from expanding into a cone.  The stream
appears to bend toward the west in the \oiii\ flux map
(Fig.~\ref{Oiiifluxfig}), and the only expanding structure that we
observe is the one small bubble.

But at larger radii, the ambient density toward the west must drop
relative to the eastern side.  While the outflow remains tightly
confined to the east, even beyond the collimated zone, certain
Fabry-Perot maps show evidence of variations in the western side of
the outer outflow cone, suggesting a more azimuthally-extended
morphology there, e.g., break-out into a less dense region. For
example, the \ha\ velocity map (Fig.~\ref{Havelfig}) reveals high
velocity clumps at radii of $\sim$1~kpc that are distributed westward
from the sharp eastern edge over almost 90\arcdeg\ in azimuth.  These
clumps are clearly associated with the outflow, and can also be seen
as the blue-shifted emission just west of the split line region in our
most distant outflow cut, Figure~\ref{conesfig}$d$.  The fact that
this gas exhibits smaller velocities in panel $c$ of this figure
illustrates that kinematical effects from disk rotation are greater
closer to the disk.  In contrast, the emission immediately east of the
split line region lies at effectively the same velocity in both panels
$c$ and $d$ of Figure~\ref{conesfig}, as this coincides with the sharp
collimating edge on that side of the outflow.  The spatial structure
of the \ha\ velocity map (like that published in \cite{H72}) also
suggests the presence of rotating disk material that has been
entrained and gradually diverted to the outflow.

% Summary:
To summarize, the inner 350~pc of the outflow constitutes a flow down
a pipe.  The outflow is collimated, presumably by ambient and
entrained disk material, and highly inclined to our line of sight,
such that the observed radial component of the flow velocity is only
$\sim$50~\kms.  Beyond 350~pc, however, the collimation weakens and
the outflow expands rapidly as a cone of emission with an opening
angle of 25\arcdeg\ and a projected front-to-back velocity separation
of approximately 300~\kms.  This expansion is preferentially toward
the west and toward the observer.  A linearly increasing intrinsic gas
velocity with an initial value of 525~\kms\ and a gradient of
0.13~\kms\ pc$^{-1}$ matches the observations well out to a radius of
a kiloparsec.

The line-splitting phenomenon indicates that the \ha-emitting
filaments are produced on the surface of the outflow cones, at the
interface between the wind and the ambient halo material.  In addition
to the minor axis velocity structure, this model explains the
increased flux within the collimated region (see
Fig.~\ref{conesfig}$a$) as a result of an elevated density with
respect to the outer expanding cone (${\cal F}\propto n_e^2$).  An
increased density in the innermost regions has also been indicated
observationally by the ``filling in'' of the line profiles at those
radii (e.g., \cite{MGDP95}).

While the inner bubble has an extremely small opening angle, we find
that an outer cone of opening angle $\sim$20\arcdeg--30\arcdeg\ fits
the data most closely.  This range agrees well with the ``small''
opening angle values found by other authors (e.g., $\sim$30\arcdeg;
\cite{MGDP95}).  Our observations do not support the ``large'' opening
angle regime (e.g., $\sim$60\arcdeg; e.g., \cite{HAM90}).

\subsubsection{Comparison of Kinematic with Hydrodynamic Models}
Striking support for the two-zone model is obtained from a comparison 
with recent hydrodynamic outflow models.  Initial two-dimensional 
simulations (e.g., \cite{TI88}) proposed an evolution of the outflow 
which included collimation of the flow at early stages by disk 
material, subsequent breakout of the confined flow along the minor 
axis, and eventual extension of the hot wind material up to a few kpc 
from the nucleus.

More recent simulations (\cite{TB93}; \cite{SBHL94}; \cite{SBHB96})
include a separate corotating halo component and are able to reproduce
much of the observed morphology through the interactions of the wind
with the disk and halo.  These simulations show that the outflow
entrains disk gas around itself, dragging the cooler, denser material
up to a couple of kiloparsecs above the plane of the galaxy (e.g., see
Fig.~6 in \cite{SBHL94}).  The regions of densest entrained disk
material, near the base of the outflow, serve to collimate the outflow
beyond the height of the disk itself.  The scale height of this
collimation in the simulations is similar to that seen in our
observations, $\sim$500~pc.  The ``fingers'' of disk material
entrained to heights above the collimated zone can be identified with
the optical emission line filaments we observe on the outer walls of
the outflow cone.  This entrained gas has also been observed at
molecular wavelengths (e.g., \cite{SC84}; \cite{NHHSHS87};
\cite{SNHGRW90}).  Within the collimated region, the simulations show
that both the wind and the confining walls are at their densest,
consistent with the observed increased levels of optical emission.
Finally, it should be noted that a recent analysis of the minor-axis
X-ray distribution implies a partially confined outflow of this gas as
well, within 1.6~kpc of the disk (\cite{BST95}).

\subsubsection{Comparison of Optical and X-ray Observations}
\label{compoptxray}
A number of authors have observed an extensive nebula of soft X-ray
emission along the minor axis of M82 (e.g., \cite{WSG84};
\cite{SPBKS89}; \cite{TOMMK90}; \cite{BST95}).  The initial
interpretation was that these are thermal X-rays, produced by the hot
($T\sim10^7$~K) gas in the outflow.  The optical line emission must
then arise from the cooler boundary layer of the wind, where the hot
gas interacts with entrained disk and ambient halo material.  Such a
hot gas would not be gravitationally bound to the galaxy and could
easily expand to the the observed radial distances of 5--6~kpc
(\cite{BST95}) in the estimated age of the starburst ($t\sim 5\times
10^7$~yr; \cite{RLTLT80}; \cite{DM93}).  Recent ASCA observations
(\cite{THAKFIIOPRN94}; \cite{ML97}) find three temperature components
in this wind, with more extended softer components, confirming that
the hot wind is cooling as it expands.

Recent hydrodynamic simulations and studies of the x-ray halo spatial
distribution have cast doubt on this interpretation, however.  Current
hydrodynamic simulations (\cite{TB93}; \cite{SBHL94}) derive gas
temperatures an order of magnitude larger than earlier estimates
(e.g., \cite{WSG84}).  This $10^8$~K gas does not emit as strongly in
the X-ray bands, producing insufficient soft and hard X-rays to
account for the observations.  Even \cite{CC85} admit that the density
falls too rapidly in their simple model to account for the x-ray
photons as thermal.  X-ray spectral studies derive a range of thermal
gas temperatures and suggest alternative emission mechanisms
(\cite{F88}; \cite{SPBKS89}; \cite{SPS96}).  Finally, correlations
between our deep \ha\ imaging and Fabry-Perot observations and
high-resolution {\it ROSAT\/} imagery support a non-thermal origin for
at least a portion of the minor-axis x-ray emission.

\placefigure{rosatfig}

Figure~\ref{rosatfig} compares the spatial distribution of soft X-rays
observed by {\it ROSAT\/} with our Fabry-Perot \ha\ flux map.  As was
evident at larger radii in the comparison with the deep \ha\ imagery
(Fig.~\ref{deephafig}), the X-rays and optical line emission are
clearly correlated.  On large scales, the minor-axis x-ray flux drops
at a radius of approximately 500~pc from the nucleus, as does the \ha\
emission.  But also on scales as small as 10 pixels (150~pc), the
x-ray and optical emission appears well correlated.  This implies that
soft X-rays are being produced in regions very close to those which
are producing \ha\ emission, a situation which is very difficult to
understand in terms of a thermal emission mechanism.

These observations lend support to the x-ray emission mechanism
suggested by several authors (e.g., \cite{CC85}; \cite{SBHL94};
\cite{SPS96}): the soft X-rays arise from shocked disk and halo
``clouds.''  This shocked gas can produce both the observed x-ray and
optical emission, accounting for the spatial correlation in
Figure~\ref{rosatfig}.  A hybrid model seems necessary in which the
higher gas temperatures and densities near the nucleus create a region
dominated by shocks at interfaces with disk and halo gas clouds, while
the cooler temperatures and lower densities at larger radii produce a
decrease in optical emission and an increase in thermally-emitted
x-rays.  A comparison of the scale lengths of the \ha\ and x-ray
emission along the minor axis confirms the more extended nature of the
x-ray component: the \ha\ surface brightness along the minor axis is
fit well by an exponential function, with a scale length of
$\sim$250~pc.  For the most distant \ha\ emission ($r\sim1$~kpc), this
exponential can be approximated by a power law of slope $-2$,
essentially the same power law exponent measured for the X-rays at a
comparable radius (\cite{BST95}).  Beyond this radius, the optical
surface brightness falls more rapidly than does the x-ray surface
brightness.

Recent detailed modeling of the x-ray emission (\cite{BST95}) suggests
a temperature at large radii of only $\sim 2 \times 10^6$~K, implying
an increasing role for thermal emission with radius.  Similarly,
hydrodynamic simulations find that the majority of the wind mass must
be accumulated near the starburst region, not from evaporating halo
clouds (\cite{SBHB96}).  It should be noted, however, that
observations with the Ginga x-ray satellite have made the startling
discovery of faint x-ray emission extending several {\it tens\/} of
kiloparsecs from M82 (\cite{TOMMK90}).  Hydrodynamic simulations have
modeled this emission as shock-excited in nature, assuming that the
outflow is much older, $\sim 5 \times 10^7$~years (\cite{TB93}).
Although this observation has yet to be confirmed, the rapid radial
decrease in the wind pressure and density could propagate the wind
shock to large distances from its starburst origins.

The true importance of shocks versus thermal emission can be estimated
from optical line diagnostics, at least in the inner regions of the
M82 outflow.  As was pointed out in Section~\ref{maps}, the
flux-weighted \niiha\ ratio from the Fabry-Perot data is highly
uniform and low in the inner kiloparsec of the outflow; values of
0.3--0.6 are typical.  However, we must be careful to use line ratios
for individual kinematic components when drawing conclusions regarding
the physics of the gas, especially when the components have been
modeled as distinct physical regions.  The \niiha\ line ratio of the
individual components reveals a similar low value across the spatial
extent of split lines, except in the inner collimated zone, where a
higher \niiha\ ratio is seen ($\sim$1.0), particularly in the
low-velocity component.  Although we were unable to resolve separate
components in the \oiii\ observations, we note that the \oiiiha\ ratio
exhibits a strong radial gradient, unlike the \niiha\ ratio.  The
\oiiiha\ ratio increases from a value of approximately 0.03 at the
center to 0.08 at a distance of $\sim$750~pc.

\placefigure{vofig}

In order to investigate the importance of shock excitation for the
optical filaments, we have compared the observed emission line ratios
from the Fabry-Perot data with the standard emission-line galaxy
diagnostic diagrams of \cite{VO87}.  Although the small number of
emission line diagnostics at our disposal limits our analysis, we can
nevertheless make a rough assessment of the influence of shocks using
the \oiiihb\ versus \niiha\ diagnostic diagram (Fig.~\ref{vofig}; from
Fig.~1 of \cite{VO87}).  Using an \hbha\ ratio of 0.25 for the outflow
gas (\cite{HAM90}), we see that the emission line ratios from the
southern wind lobe of M82 rest in the region of the diagram for
starburst galaxies, as expected.  The ratios are comparable to those
for most cooler \hii\ regions and \hii\ region models.  This
immediately suggests an emission mechanism such as photoionization for
the filaments, particularly near the nucleus, where the \oiiiha\ ratio
is lower.  As we move out in radius, however, shocks appear to become
more important as an excitation mechanism, as the increasing \oiiiha\
ratio drives the locus in Figure~\ref{vofig} toward the region for
non-thermally powered AGN.

\placefigure{dsfig}

In order to more directly interpret our emission line fluxes in light
of a shock mechanism, we have also compared the observed line ratios
with a recent set of high-velocity shock models (\cite{DS95}).  These
models have been computed for shocks in the velocity range of
150--500~\kms; the deprojected gas velocity of the wind in M82 is
estimated to be at the upper end of this range.  Again using an \hbha\
ratio of 0.25 for the outflow gas (\cite{HAM90}), a comparison with
the \oiiihb\ versus \niiha\ diagnostic diagram (Fig.~\ref{dsfig}; from
Fig.~2$b$ of \cite{DS95}) shows that it is unlikely that the observed
emission line flux from the inner outflow filaments arises entirely
from shocks.  There is simply not enough \oiii\ emission observed in
the inner kiloparsec of the M82 outflow.  However, the increasing
\oiiiha\ ratio with distance from the nucleus suggests that shocks
probably become important at larger radii, just as suggested by the
observational diagnostic diagrams (\cite{VO87}).  Longslit optical
observations have also reached the conclusion that the line ratios
become more shock-like with increasing distance from the starburst
(e.g., \cite{HAM90}), although this has often not included analysis of
individual velocity components.

Additional support for a photoionization mechanism for the inner
optical filaments is provided by studies of the diffuse ionized medium
(DIM) in NGC~891.  In that galaxy, which has no outflow or other
obvious sources of shock ionization, photoionization models have been
used to understand the variation of line ratios with height above the
disk plane.  These models show that the \niiha\ ratio gradually
decreases as the ionization parameter (the ratio of ionizing photons
to gas density) increases.  In contrast, the \oiiiha\ ratio should
increase rapidly with ionization parameter (\cite{S92}).  Regardless
of the presence of shocks then, the low value of \niiha\ and the
gradually increasing value of \oiiiha\ in the innermost filaments of
M82 can be understood as the result of a gradual drop in filament
density, relative to the number of ionizing photons from the
starburst.  In the outer filaments, however, the \niiha\ ratio begins
to increase, presumably as a result of dilution of the radiation field
in the expanding uncollimated bubble, as well as perhaps an increasing
influence of shocks.

Recent studies of the influence of halo dust on these line ratio
trends in NGC~891 (\cite{FBDG96}) point out that the \niiha\ ratio may
appear artificially low near the disk due to dilution by scattered
radiation from disk \hii\ regions. This suggests that the low \niiha\
ratios for the inner filaments in M82 may be due in part to higher
dust densities in the inner halo, scattering disk radiation from the
nuclear starburst.  This proposition corresponds well with the high
levels of polarization detected from the filaments and the exponential
nature of the observed halo, although the observed polarization levels
in M82 ($\sim$10--15\%) are much higher than those modeled in NGC~891
($\sim$1--2\%).

Based upon these comparisons and our geometric models, we propose that
the optical emission from the inner kiloparsec of the M82 filament
network is, at least partially, due to photoionization of the sides of
the cavity created by the outflow.  The hot gas in the wind itself
would be quite transparent to the UV ionizing photons from the
starburst region, allowing the entrained disk and halo gas to be
illuminated directly.  The tipped geometry of the outflow cones
probably places the systemic side of each cone more directly in the
path of the photoionizing radiation from the central starburst,
explaining the higher fluxes seen in the low-velocity components of
the wind.  However, small regions of higher \niiha\ ratios in the
individual velocity components suggest that a complex combination of
shock and photoionization is probably required in the violent
collimated zone, where the disk gas is being entrained and drawn
upward by the hot wind just as it leaves the luminous starburst
region.  Although our small field of view and sensitivity limits
restrict our analysis of the more extended optical filaments, we
confirm a trend toward more shock-like line emission in the outer
regions of the outflow.

\subsubsection{Comparison with UV Observations}
The inner regions of the outflow in M82 have been detected in UV
images taken with the {\it Ultraviolet Imaging Telescope\/} (UIT;
\cite{Se92}).  Figure~\ref{uvfig} compares our deep \ha\ image with
the UV image observed in the 2490\AA\ band by the {\it UIT}.  The
southern outflow is clearly evident in ultraviolet emission, out to at
least a kiloparsec.  The northern outflow is only barely visible in
the {\it UIT\/} data, presumably because of the high disk extinction
at ultraviolet wavelengths.  The projected morphology of the southern
UV outflow is similar to that at \ha\ wavelengths, consisting of a
relatively smooth, broad fan of emission.  The ratio of the \ha\ and
UV surface brightnesses (each in units of ergs s$^{-1}$ cm$^{-2}$
arcsec$^{-2}$ \AA$^{-1}$) in the outflow region is $\sim$8.8
(\cite{He96}).

\placefigure{uvfig}

Previous analyses of the {\it UIT\/} data (\cite{MOLNRSSS91};
\cite{H93}), as well as earlier balloon observations (\cite{CRBGH90};
\cite{BGHRB90}) have interpreted the minor-axis ultraviolet emission
in M82 in terms of dust scattering of photons from the nuclear
starburst by particles in the outflow.  It has been suggested that
filamentary structures can be seen in the {\it UIT\/} data
(\cite{MOLNRSSS91}), presumably because the higher relative densities
in the filaments enhance scattering in those regions.  The spatial
distribution of the UV emission does not correlate well with the \ha,
X-ray, radio continuum, or IR morphologies, although this may be due
to variations in opacity (\cite{CRBGH90}).  One difficulty with the
scattering picture for the production of the southern UV outflow is
the surprising non-detection of emission in the {\it UIT\/} 1600\AA\
far-ultraviolet (FUV) band (\cite{H93}).  The naive expectation for
Rayleigh scattering would be to expect increased scattered flux at
shorter wavelengths, even in the presence of substantial extinction
along the line of sight.  Ultraviolet observations of reflection
nebula (e.g., \cite{CBGWB95}) and starburst galaxies (\cite{CKS94})
derive extinction curves that confirm this expectation.

An alternative explanation for the UV light is ``two photon''
continuum emission from ionized gas within the optical filaments.
This emission arises from the spontaneous two-photon decay of the
2~$^2S$ level of \hi\ and is commonly seen in AGN and regions of
low-density ionized gas.  The spectral distribution of two-photon
emission is symmetric, in photons per frequency interval, about a peak
at 2431\AA\ (\cite{O89}), very close to the {\it UIT\/} 2490\AA\ band.
Unlike the scattering function, two-photon flux decreases
($\sim\lambda^{-1}$) toward shorter wavelengths (see also
\cite{He96}).  Preliminary analysis of the \sii\
$\lambda\lambda$6719,6731 lines in optical spectra of the outflow from
the {\it Keck\/} telescope confirm previous studies (e.g.,
\cite{RC80}; \cite{HSGH84}; \cite{HAM90}) that the inner filament
densities rarely exceed 500 cm$^{-3}$.  As this is significantly below
the critical density at which collisional effects influence the
intensity of two-photon emission ($n_e\sim10^{4}$~cm$^{-3}$;
\cite{O89}), we may anticipate the filamentary nature of the outflow
in the UV imagery.

The observed \hbha\ ratio of 0.25 for the outflow gas (\cite{HAM90})
implies an \hb/UV ratio of $\sim$2.2.  This is approximately a factor
of ten greater than the ratio of \hb\ to two-photon flux predicted by
photoionization models ($\sim$0.1; \cite{DBS82}; \cite{D97}),
suggesting the presence of 2.5 mag more extinction at 2500\AA.  The
observed extinction law for starburst galaxies (\cite{CKS94}; Fig.~5),
combined with the observed \ha\ extinction factor of 1.8 in the
outflow (\cite{HAM90}) indicates that this level of obscuration is
entirely reasonable.  Shock-induced two-photon emission is almost
certainly ruled out, as it would require extinction by a factor of
$\sim4$ to match the observed UV intensity, well above the value
observed in the outflow regions.  Spectral observations near the
Balmer limit, $\sim$3650\AA, where the two-photon continuum emission
is enhanced relative to line emission, could be useful for determining
the importance of this mechanism.

\section{Conclusions}
\label{conclusions}
It is widely recognized that M82 is the prototype of galactic wind
systems. Most models have concentrated on explaining measurements
taken along isolated position angles close to the galaxy's minor axis.
However, our new study has shown that the filamentary system
associated with the outflow is highly complex. Our simplest kinematic
model requires at least two discrete structures for each side of the
galaxy. The filaments are distributed in a network over these
surfaces.  More surprisingly, the axes of the outflows on either side
of the disk are aligned neither with each other nor with the spin axis
of the disk. The gas excitation suggests that stellar ionization
dominates close to the disk with increased dominance of shock
excitation further along the outflow. The clearest signature of the
biconic outflow geometry is seen in the line ratio maps which suggest
that radiation is escaping from the disk along a channel excavated by
the hot rarefied wind. We find evidence for a smooth line-emitting
halo which we associate with the linearly polarized halo seen in
broadband studies.  We have observed a warm ionized medium throughout
the inclined spiral disk.

In our view, the most pressing issue is an explanation of the
emission-line spectropolarimetry. We strongly urge that these
observations are repeated to deeper levels, at least to the point
where the diffuse line-emitting halo is detected. Observations which
{\it only\/} detect the bright filaments cannot remove the
contribution from the halo along the line of sight.

The disk-halo interaction in galaxies is a fundamental topic in its
own right (\cite{B90}). We suggest that M82 may provide the best
observational constraints on fountain flows (e.g., \cite{SB91}) once
the origin of the filaments is fully understood. What happens to the
entrained material that is lifted into the galactic halo?  What is the
relationship between the relativistic electron halo, the wind
filaments, and the diffuse line-emitting halo?  We suspect that
answers can only come from future observations of x-ray, radio and
millimeter emission at a resolution and sensitivity comparable to the
present study.

\acknowledgments

Funding for this research was provided by the office of the Dean of
Natural Sciences at Rice University, the Texas Space Grant Consortium,
and the Sigma Xi Research Society.  Additional funding was provided by
AURA/STScI (grant GO-4382.01.92A), the National Science Foundation
(grant AST 88-18900), the William F. Marlar Foundation, the National
Optical Astronomy Observatories (NOAO), and Mr.\ and Mrs.\ William
Gordon.  This research was performed in partial fulfillment of the
Ph.D. degree at Rice University. JBH acknowledges a Fullam Award from
the Dudley Observatory.

We thank Drs.\ Reginald J. Dufour, Patrick M. Hartigan, Jon C.
Weisheit, C. R. O'Dell, and Sylvain Veilleux for enlightening
discussions of the material contained herein.  Special thanks is due
Drs.\ Joel Bregman and Brent Tully for providing the {\it ROSAT\/} and
deep \ha\ imagery, respectively, and Drs.\ Geoff Bicknell and Michael
Dopita for their assistance with the stellar wind equations and
high-velocity shock models, respectively.

This research has made use of the NASA/IPAC Extragalactic Database
(NED) which is operated by the Jet Propulsion Laboratory, Caltech,
under contract with the National Aeronautics and Space Administration.
The Astro\-physics Science Infor\-mation and Abstract Service (ASIAS),
administered by the Astrophysics Data System (ADS), was also used.
Color versions of several figures in this paper are available from the
authors.

%
% References
%

%
% Figure captions
%
\newpage

\figcaption[fig01.eps]{Flux maps of M82 in the emission lines of
  ($a.$) \ha\ 6563\AA\ and ($b.$) \nii\ 6583\AA, in units of ergs
  cm$^{-2}$ sec$^{-1}$, log-scaled between $-15.5$ and $-13.0$.  The
  upper maps represent the total flux in the line at each pixel, while
  the lower panels represent the flux in the high- (HVC) and low-
  (LVC) velocity components of the line, in the regions where the line
  is split (designated by an outline in the upper panel).  Splitting
  of the \nii\ emission line is only observed south of the disk.
\label{HaNiifluxfig}}

\figcaption[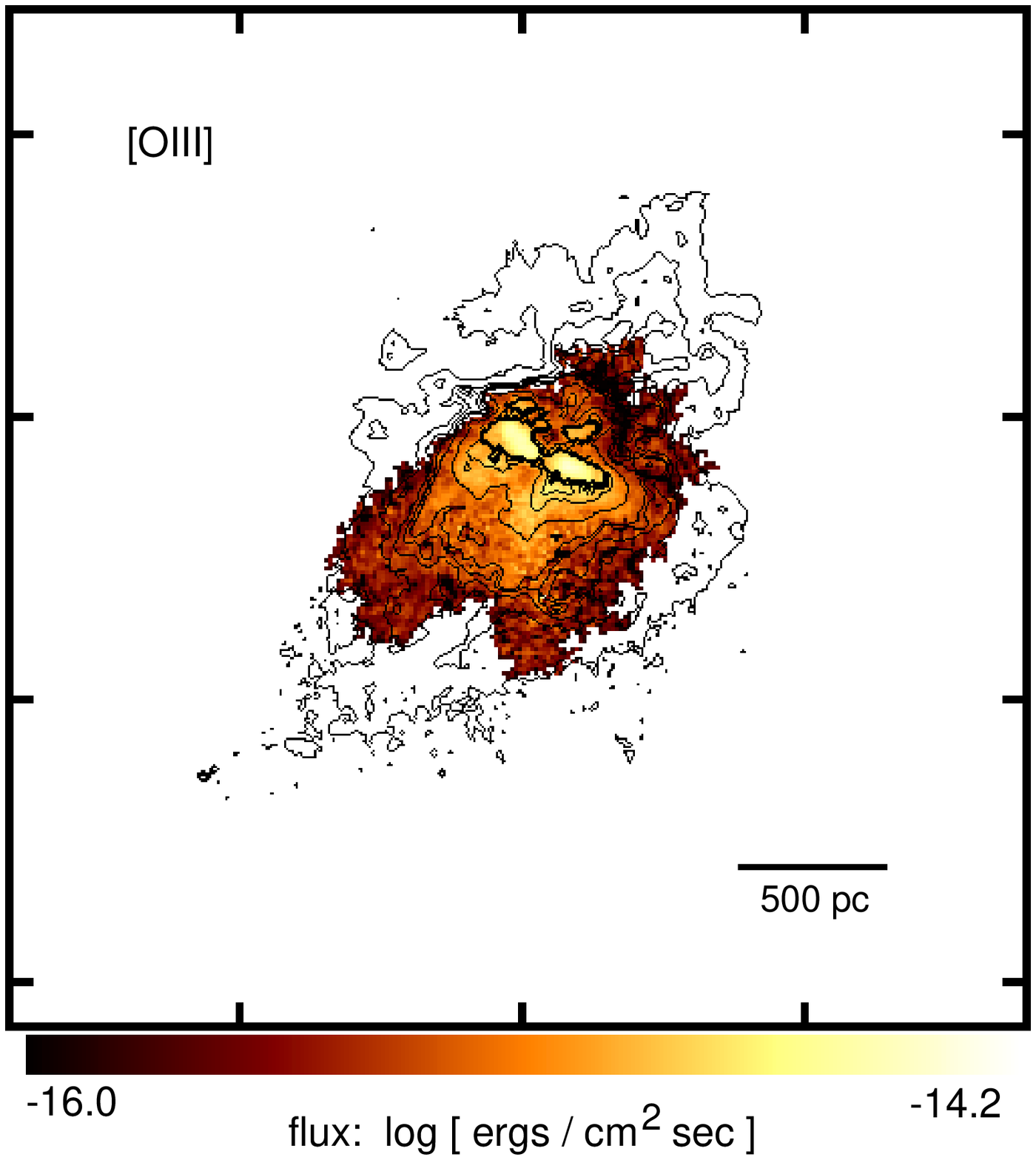]{Flux map of M82 in the emission line
  of \oiii\ 5007\AA, in units of ergs cm$^{-2}$ sec$^{-1}$, log-scaled
  between $-16.0$ and $-14.2$.  The \oiii\ emission line profiles were
  fit with single components throughout the field.  A contour map of
  the \ha\ flux has been superimposed.
\label{Oiiifluxfig}}

\figcaption[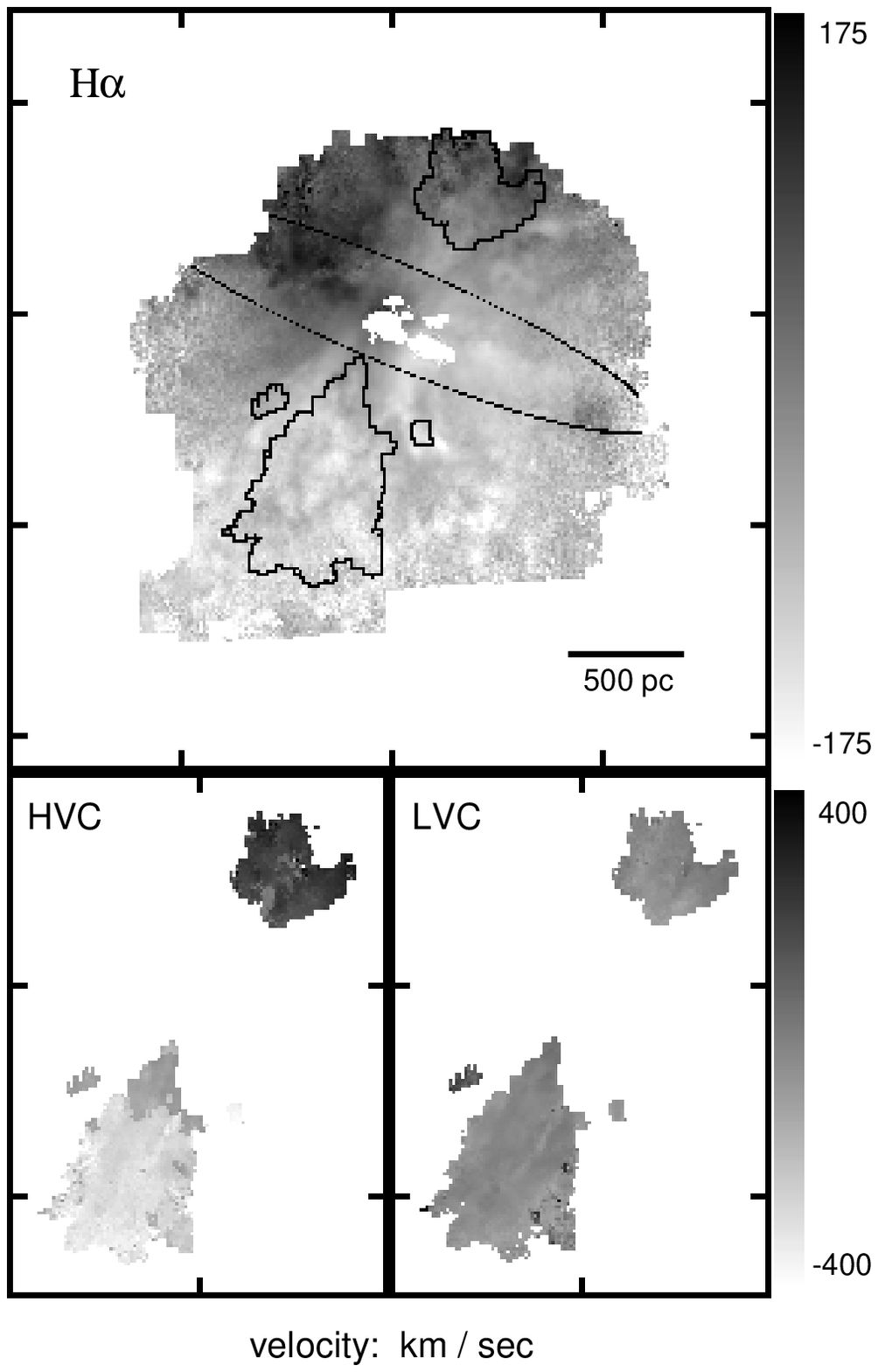]{Velocity map of M82 in the emission line of
  \ha\ 6563\AA, in units of \kms.  The upper map represents the total
  velocity of the \ha-emitting gas at each pixel.  In regions of
  multiple velocity components, the component velocities have been
  combined with flux weighting to yield an ``average'' velocity.  An
  ellipse centered on the 2.2~\micron\ nucleus (\cite{WGT92}) has been
  superimposed on the image.  The lower panels represent the
  velocities in the high- (HVC) and low- (LVC) velocity components of
  the line, in the regions where the line is split (designated by an
  outline in the upper panel).  Note that the velocity range is much
  greater for these component panels than for the total velocity map.
\label{Havelfig}}

\figcaption[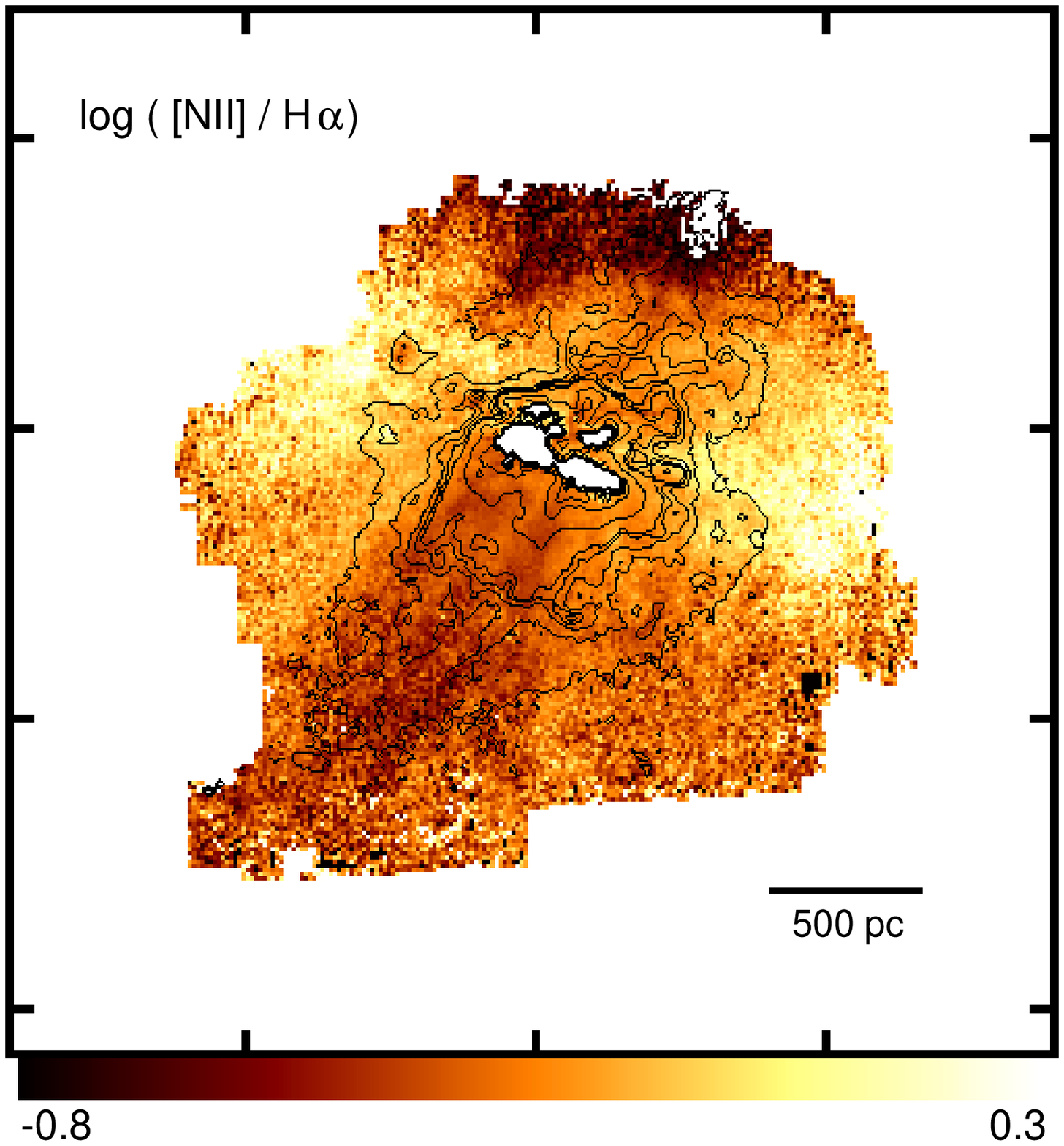]{Flux ratio map of M82 for the ratio of the
  flux in the emission line of \nii\ 6583\AA\ to the flux in the
  emission line of \ha\ 6563\AA, log-scaled between $-0.8$ and $+0.3$.
  A contour map of the \ha\ flux has been superimposed.
\label{NiiHalgfig}}

\figcaption[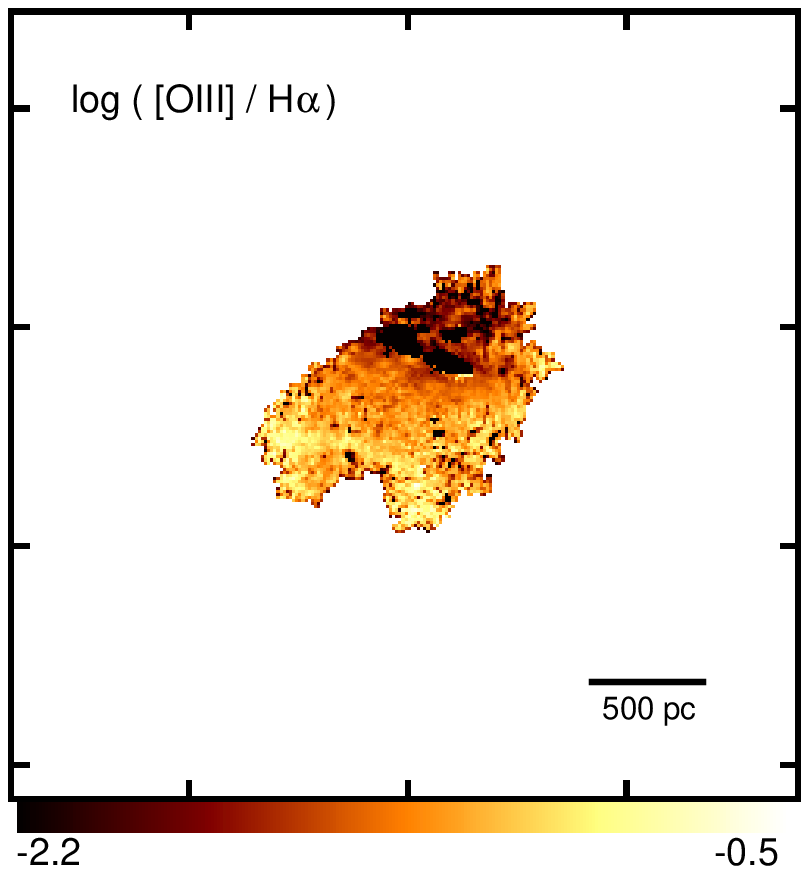]{Flux ratio map of M82 for the ratio of the
  flux in the emission line of \oiii\ 5007\AA\ to the flux in the
  emission line of \ha\ 6563\AA, log-scaled between $-2.2$ and $-0.5$.
\label{OiiiHalgfig}}

\figcaption[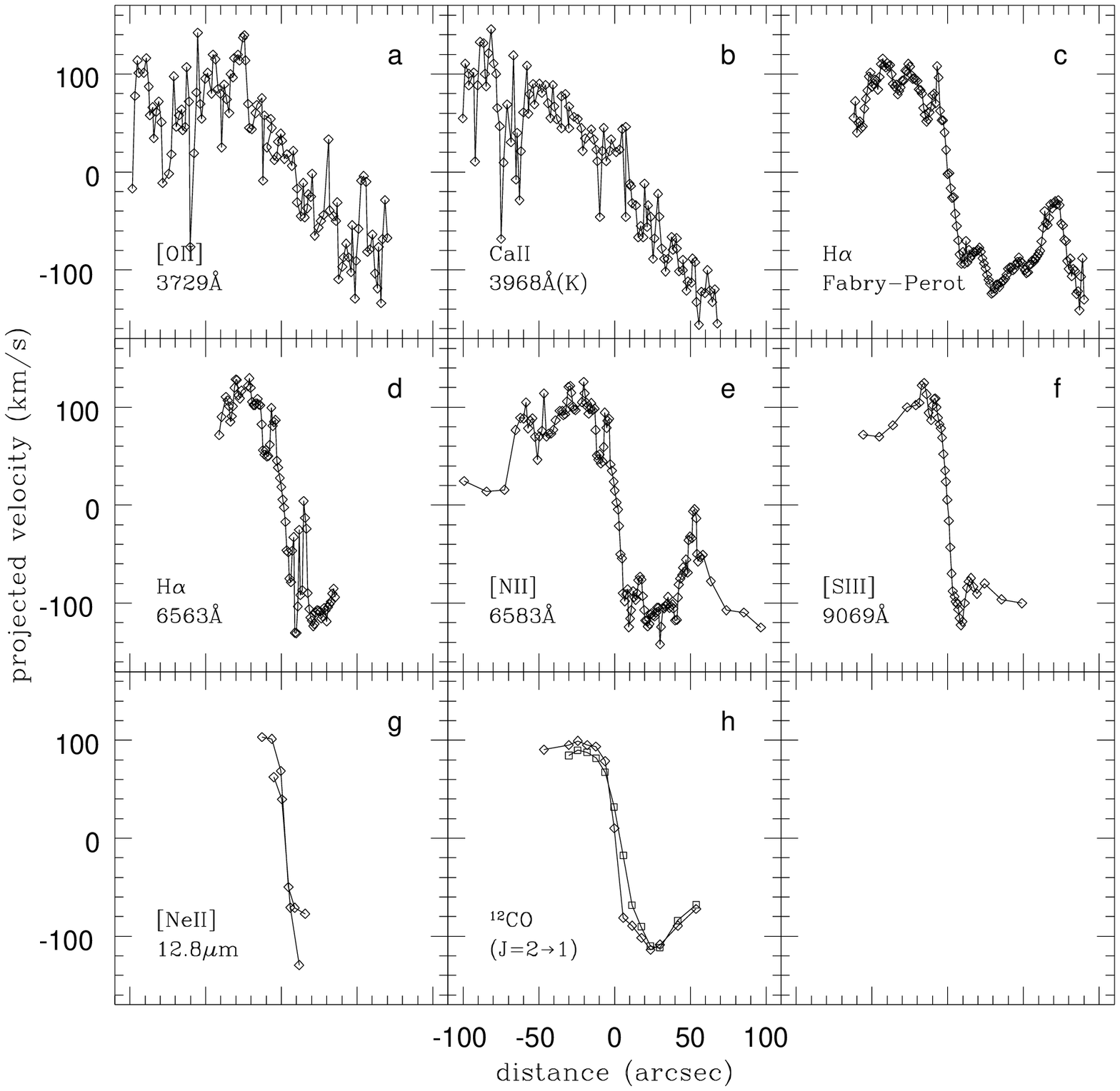]{Rotation curves from the major axis of M82,
  in order of increasing wavelength.  Literature sources include:
  $a.$--$b.$ \cite{MCGD93}, $c.$ this paper, $d.$--$e.$ \cite{CMG91},
  $f.$ \cite{MCGD93}, $g.$ \cite{BLBT78}, $h.$ \cite{LNSWRK90}.  A
  systemic velocity of 208.7~\kms\ has been subtracted from the
  Fabry-Perot data.
\label{rotcurvefig}}

\figcaption[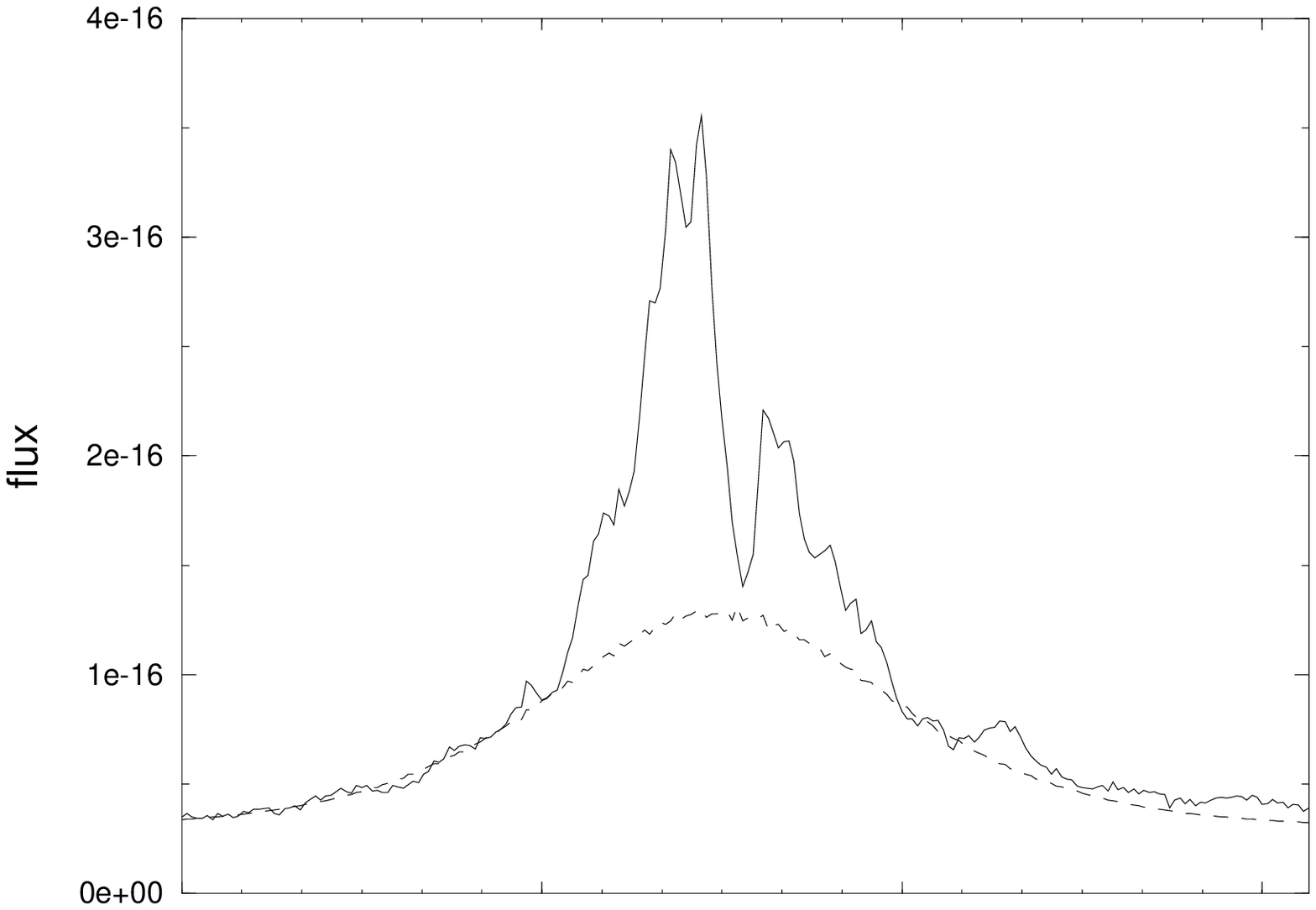]{A profile of the peak \ha\ emission along a
  band parallel to and approximately 725~pc south of the major axis of
  M82, illustrating the bright filaments and underlying diffuse halo
  component.  The dashed line represents a cut through an exponential
  halo model.  Tick marks on the spatial axis are spaced at 1\arcmin\
  intervals.
\label{exphalofig}}

\figcaption[fig08.eps]{A deep \ha\ image of M82.  Filamentary
  structure is observed along the minor axis of the galaxy across the
  entire 6\arcmin\ field.  A contour map of the {\it ROSAT\/} HRI
  image has been superimposed; the contours represent 0.5 to 4.0
  events per pixel, in thirteen log-spaced increments.  The x-ray map
  has been smoothed with a Gaussian of FWHM $\sim$12\arcsec.  Tick
  marks are spaced at 1\arcmin\ intervals.
\label{deephafig}}

\figcaption[fig09.eps]{Mosaic of 16 monochromatic frames from the
  \ha\ Fabry-Perot data cube.  The field of view is 3\farcm 5.  The
  frames are separated by $\sim$45~\kms, where the frame marked with
  an asterisk corresponds roughly to the systemic velocity of the
  galaxy.  The circular region in the southeast corner is a masked
  ghost from the star in the southwest corner.
\label{Hamosaicfig}}

\figcaption[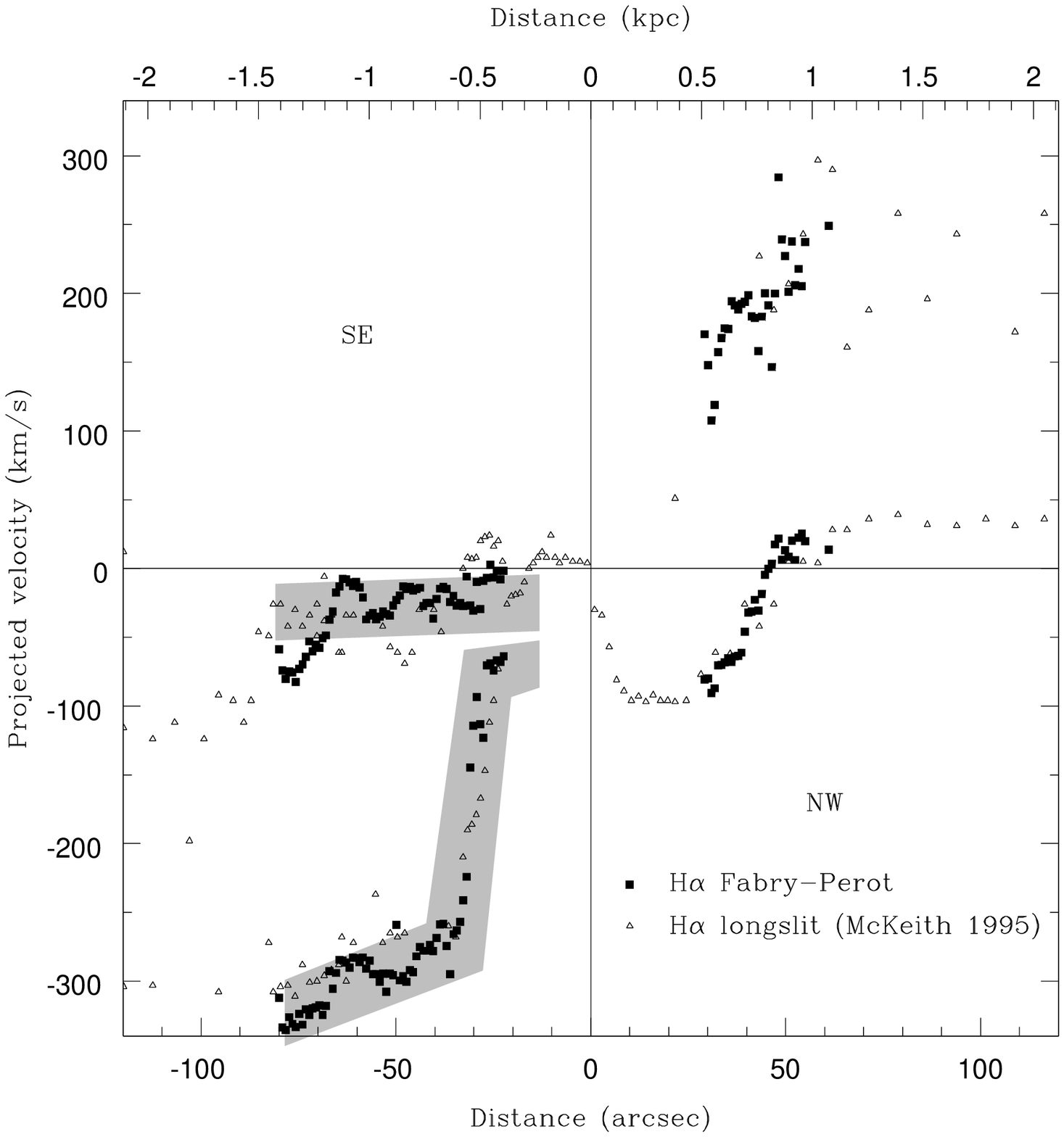]{The radial velocity profile from fits to the
  \ha\ line components along the axis of the outflow (position angle
  of 150\arcdeg), from both the Fabry-Perot data and the longslit
  observations of \cite{MGDP95}.  A systemic velocity of 200\kms\ has
  been subtracted from all values.  The Fabry-Perot points have been
  interpolated over a band 7\arcsec\ wide through the 2.2~\micron\ 
  nucleus.  Dual \ha\ components are resolved in both the north and
  south outflow lobes.  The shaded region represents a velocity cut
  along the axis of a two-cone Monte-Carlo simulation.
\label{windvelfig}}

\figcaption[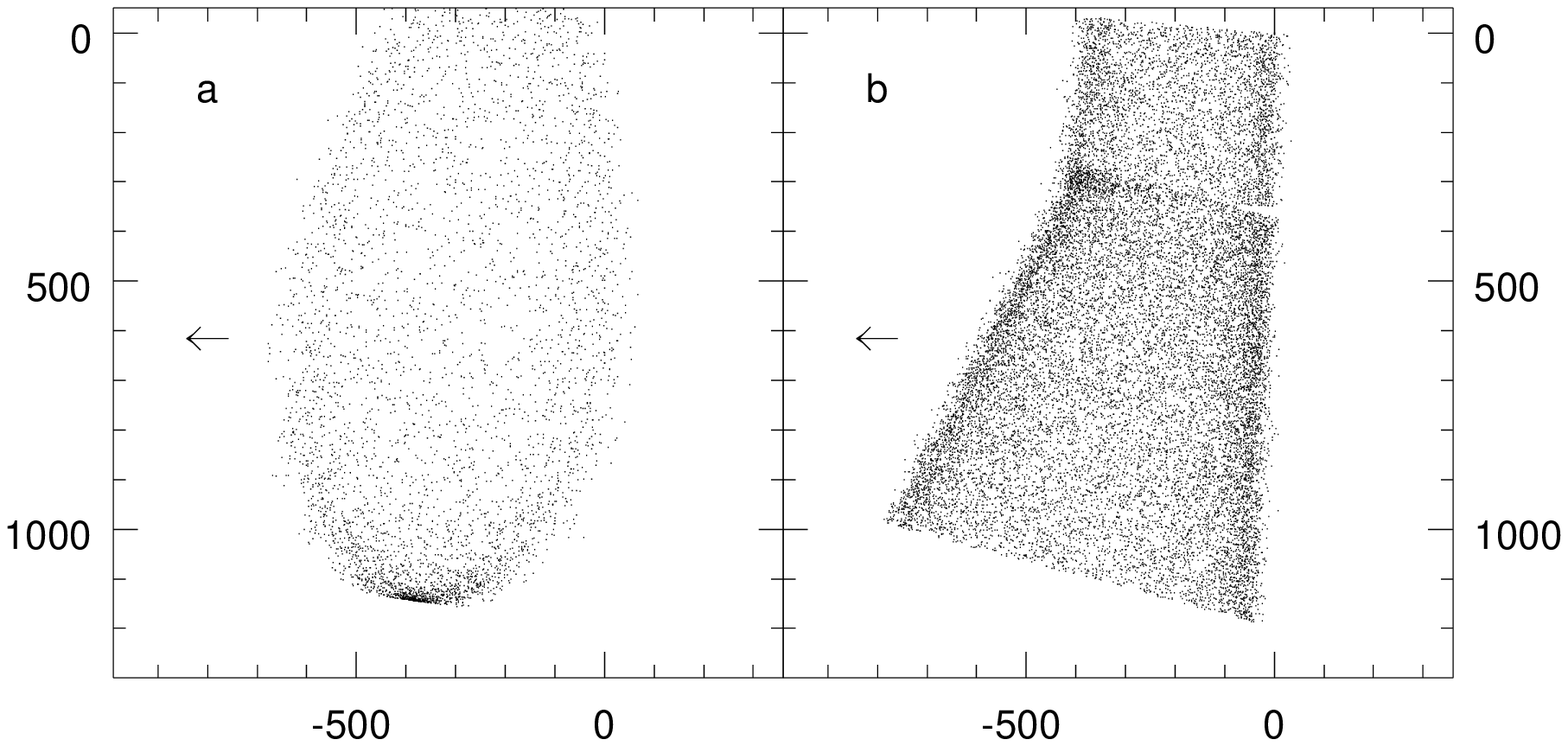]{Y-Z cuts along the axes of two Monte-Carlo
  models of the southern outflow in M82: $a.$ a single truncated
  bubble and $b.$ a pair of inclined cones.  Dimensions are given in
  parsecs; arrows denote the direction toward the observer.
\label{modelsfig}}

\figcaption[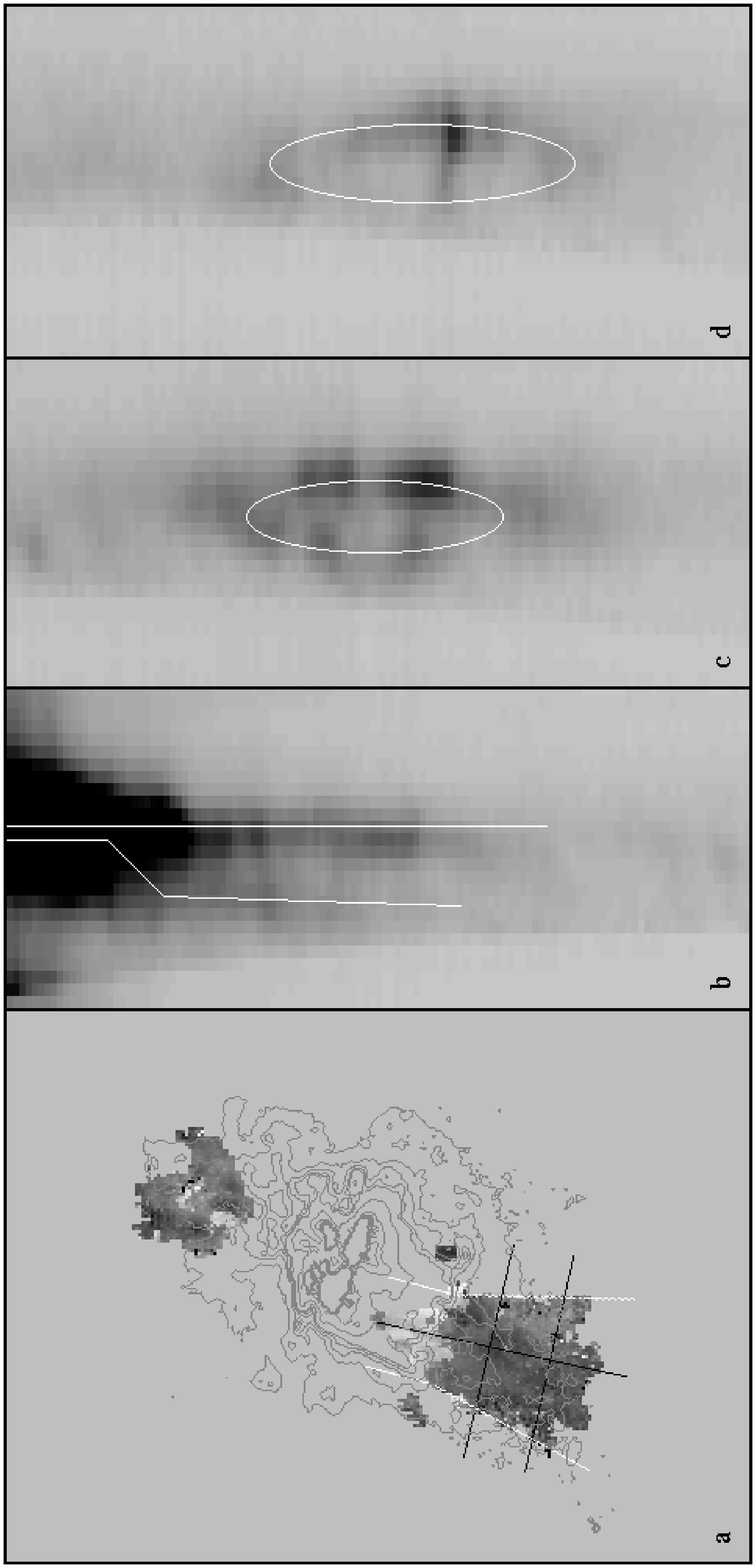]{Comparison of the observed \ha\ emission and a
  two-cone Monte-Carlo simulation of the southern outflow in M82.
  Panel $a$ compares the model with the spatial distribution of \ha\
  flux (contour map) and the regions of split \ha\ lines (grayscale).
  Panels $b$-$d$ are two-dimensional spectra extracted along the axes
  drawn in panel $a$, and illustrate the multiple velocity components.
  The spectrum in panel $b$ has been extracted along the axis of the
  cones, while panels $c$ and $d$ have been extracted perpendicular to
  the cone axis, at distances of 650 and 910~pc from the nucleus.
  Panel $b$ is oriented with North at the top; panels $c$ and $d$ are
  oriented with West at the top.
\label{conesfig}}

\figcaption[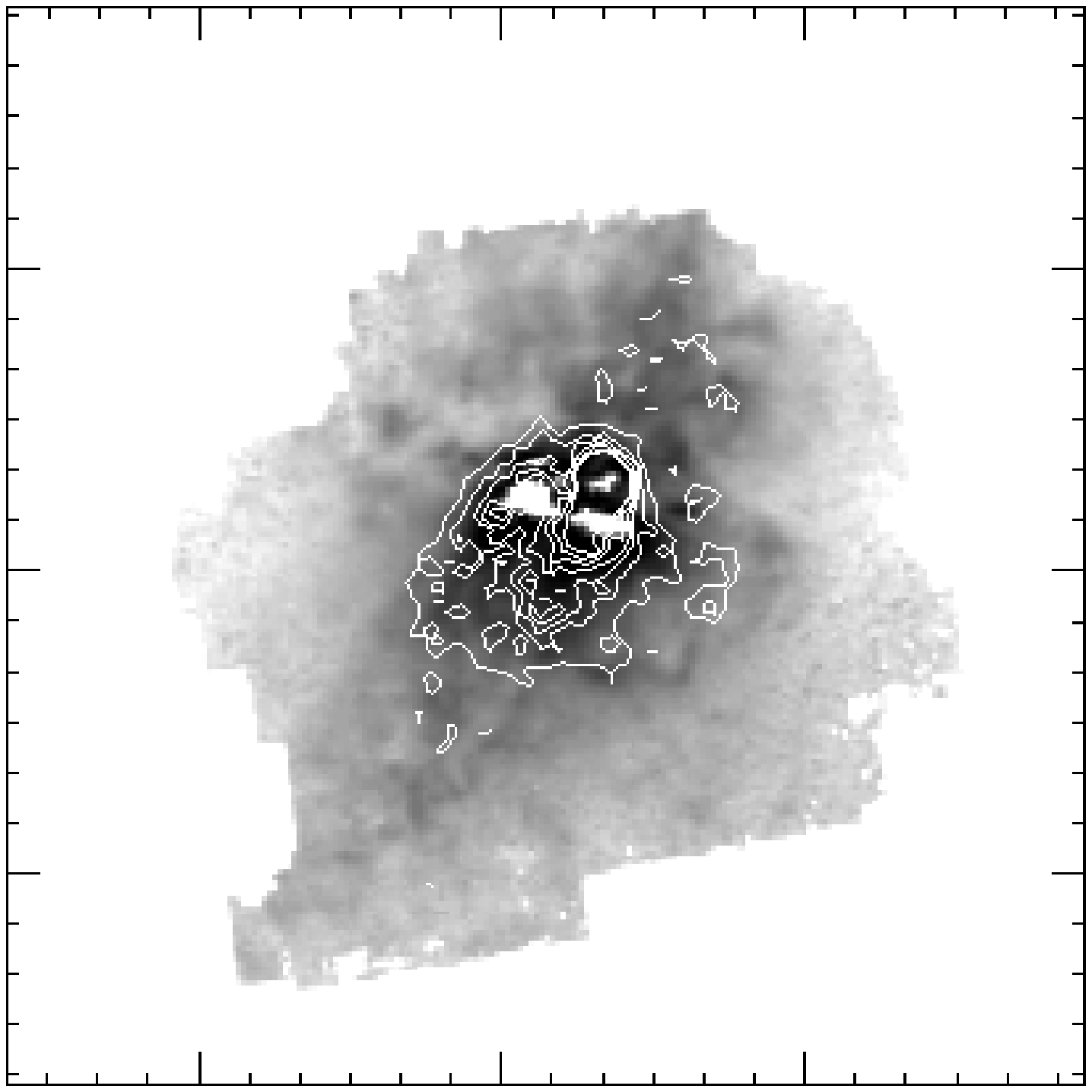]{The \ha\ flux map from the optical Fabry-Perot
  data set overlaid with a contour map of the {\it ROSAT\/} HRI x-ray
  image.  Tick marks are spaced at 1\arcmin\ intervals.  The \ha\ 
  image is displayed logarithmically between $10^{-15.5}$ and
  $10^{-13.0}$ ergs cm$^{-2}$ sec$^{-1}$, while the x-ray contours
  represent 0 to 5 events per pixel, in 7 equally-spaced increments.
  The x-ray map has been smoothed with a Gaussian of FWHM
  $\sim$2\arcsec.
\label{rosatfig}}

\figcaption[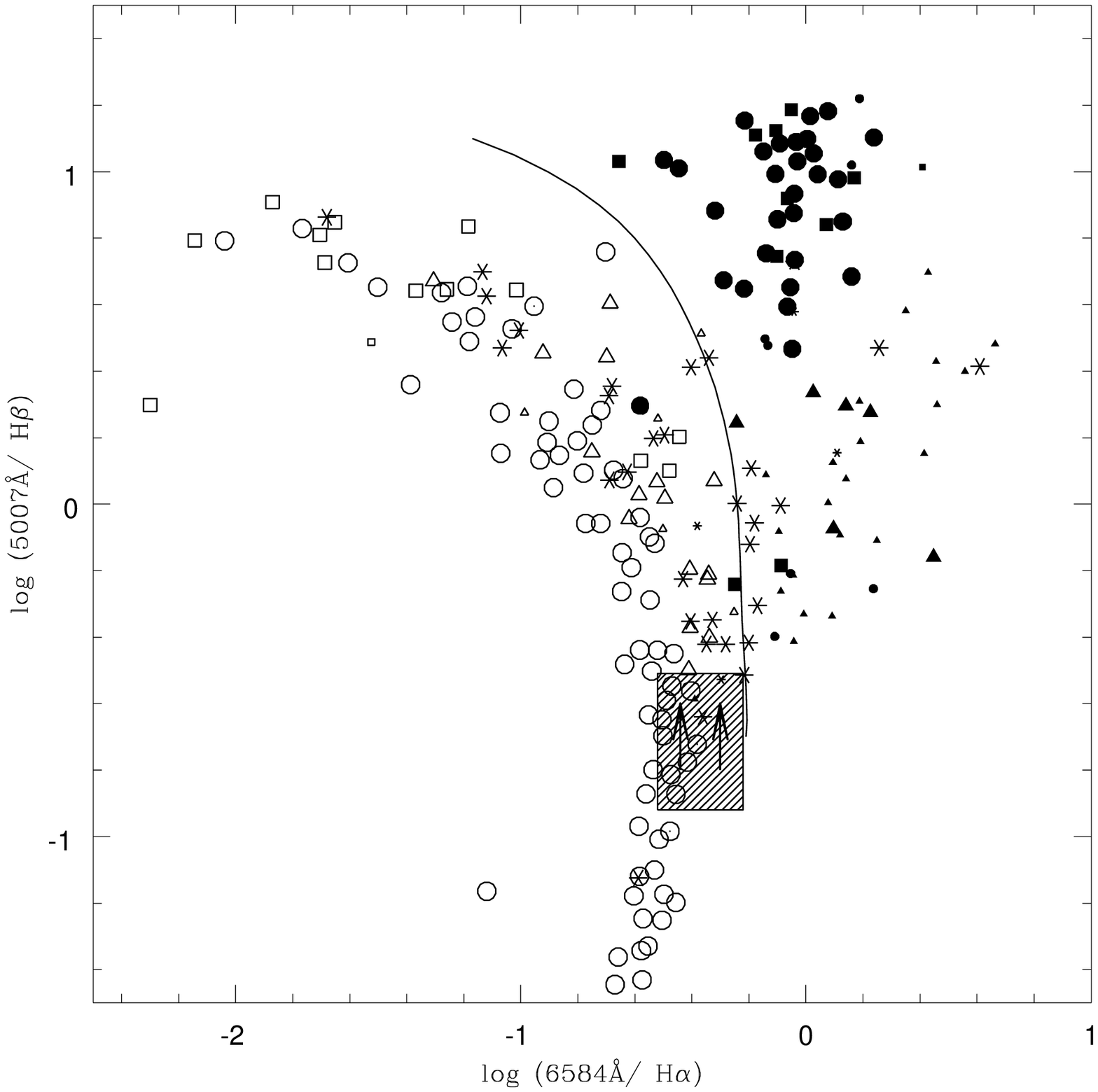]{Reddening-corrected \oiiihb\ versus \niiha\ for a
  selection of \hii\ regions (circles), \hii\ region models (lines),
  starburst galaxies (triangles), and AGN (filled polygons).  The
  position of the outflow gas in the southern lobe of the M82 wind is
  represented by the shaded region.  The arrows are oriented in the
  direction of increasing distance from the nucleus and represent a
  radial extent of approximately 1~kpc.  (Plot adapted from
  \cite{VO87}.)
\label{vofig}}

\figcaption[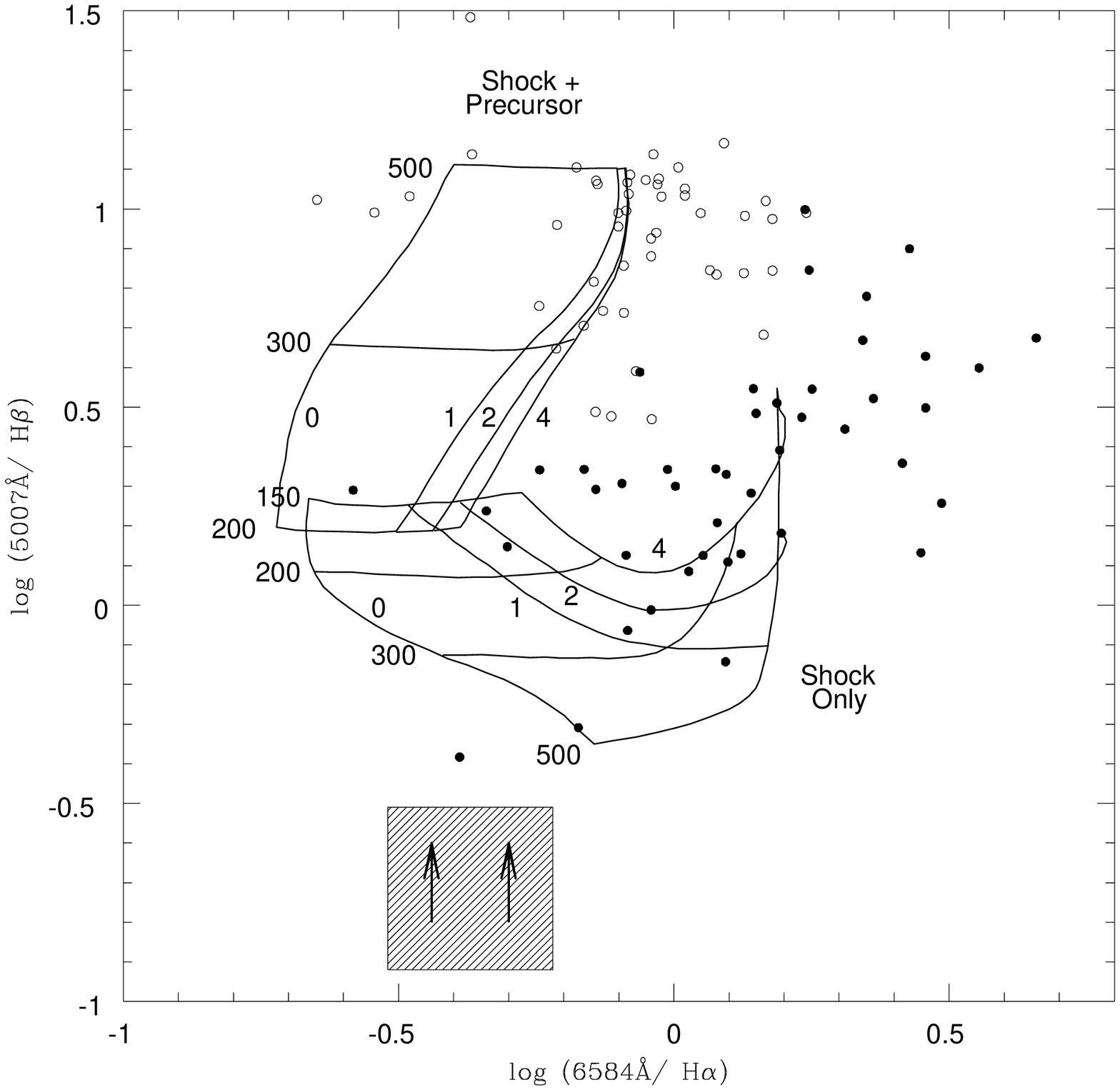]{Reddening-corrected \oiiihb\ versus \niiha\ grids
  for a selection of high-velocity shock models.  Observations of
  Seyfert galaxies are shown as open circles; LINERs are shown as
  filled circles.  The grid labeled 'Shock Only' includes only the
  emission from the shock, while the grid labeled 'Shock + Precursor'
  includes the contribution of preshock ionization.  Emission
  characteristics of shocks with velocities of 150--500~\protect\kms\ 
  are computed.  The position of the outflow gas in the southern lobe
  of the M82 wind is represented by the shaded region.  The arrows
  ares oriented in the direction of increasing distance from the
  nucleus and represents a radial extent of approximately 1~kpc.
  (Plot adapted from \cite{DS95}.)
\label{dsfig}}

\figcaption[fig16.eps]{A deep \ha\ image overlaid with a contour map of
  the 2490\AA\ ultraviolet emission observed by the {\it UIT}.  Tick
  marks are spaced at 1\arcmin\ intervals.
\label{uvfig}}

%
% Tables
%
\clearpage
\begin{deluxetable}{ll}
  \tablecaption{Useful observational values for the galaxy M82.  All
    values are taken from \cite{ddCBPF91}, except where otherwise
    noted.
\label{m82statstab}}
\small \tablewidth{5in} \startdata \nl \hline
Designations         & M82, NGC~3034, UGC~5322, ARP~337, 3C~231          \\
Position (2000.0)    & $\alpha =$ 09\fh\ 55\fm\ 54\fs0;\ $\delta =$ 69\arcdeg\ 40\arcmin\ 57\arcsec \\
Galaxy type          & I0, Irr II                                        \\
Dimensions           & 11\farcm 2 $\times$ 4\farcm 3 \quad (10.6 $\times$ 4.1~kpc) \\
Position angle       & 65\arcdeg                                         \\
Disk inclination     & 81\fdg 5 \tablenotemark{a}                        \\
Radial velocity      & $203 \pm 4$ \kms                                  \\
Distance             & $3.63 \pm 0.34$ Mpc \tablenotemark{b}             \\
Mass                 & $2.7 \times 10^{10}$ M$_{\sun}$ \tablenotemark{a} \\
\\
Total magnitude      & B = $9.30 \pm 0.09$                               \\
Galactic extinction  & A$_{\rm B}$ = 0.13                                \\
Total colors         & B - V = $0.89 \pm 0.01$                           \\
& U - B = $0.31 \pm 0.03$ \\
Radio luminosity     & L$_{\rm radio}$ = $10^{39}$ ergs s$^{-1}$ \tablenotemark{c} \\
Infrared luminosity  & L$_{\rm IR}$ = $10^{44}$ ergs s$^{-1}$ \tablenotemark{c} \\
H$\alpha$ luminosity & L$_{{\rm H}\alpha}$ = $2 \times 10^{41}$ ergs s$^{-1}$ \tablenotemark{d} \\
X-ray luminosity     & L$_{\rm X-ray}$ = $2 \times 10^{40}$ ergs s$^{-1}$ \tablenotemark{c} \\
\enddata
\tablenotetext{a}{\cite{LS63}}
\tablenotetext{b}{\cite{FHMMLSKTFFGHHHI94}.  For comparison purposes,
  we will use the standard value in the literature of $3.25 \pm
  0.20$~Mpc (\cite{TS68}) in this paper.}
\tablenotetext{c}{\cite{WSG84}}
\tablenotetext{d}{\cite{HAM90}}
\end{deluxetable}

\clearpage
\begin{deluxetable}{llllll}
  \tablecaption{Theoretical and observational characteristics of the
    Fabry-Perot instruments.
\label{HIFItab}}
\small \tablehead{
\colhead{} & \colhead{} & \multicolumn{2}{c}{\ha+\nii\ Observations} & \multicolumn{2}{c}{\oiii\ Observations} \\
\colhead{Constant} & \colhead{} & \colhead{theoretical} & \colhead{experimental} & \colhead{theoretical} & \colhead{experimental}}
\startdata

\multicolumn{2}{l}{Telescope/Instrument:} & \multicolumn{2}{l}{CFHT +
  HIFI} & \multicolumn{2}{l}{UH88 + HIFI} \nl
\hline
$D_{tel}$ & telescope diameter & 3.6 m & \nodata & 2.24 m & \nodata \nl
$f_{cam}$ & camera focal length & 90 mm & \nodata & 50 mm & \nodata \nl
\nl

\multicolumn{2}{l}{Detector:} & \multicolumn{2}{l}{IfA CCD + VISACOM \tablenotemark{a}}
& \multicolumn{2}{l}{CCD} \nl
\hline
$p_\mu$ & pixel size & 30 \micron & \nodata & 24 \micron & \nodata \nl
$p^{\prime\prime}$ & plate scale & 0\farcs 86 pix$^{-1}$ & \nodata & 0\farcs 846 pix$^{-1}$ & \nodata \nl
$n_p$ & CCD size & 256$\times$256 pix & \nodata & 432$\times$432 pix &
\nodata \nl
$t$ & exposure time & 450 sec & \nodata & 480 sec & \nodata \nl
\nl

\multicolumn{2}{l}{Etalon:} & \multicolumn{2}{l}{CFHT3} & \multicolumn{2}{l}{SV1} \nl
\hline
$\Delta\lambda$ & free spectral range (FSR) & 85 \AA & 87.9 \AA & 61
\AA & 55 \AA \nl
$N_R$ & reflective finesse & 60 & 45-50 & 60 & \nodata \nl
$n$ & order of interference & 75 & 72 & 91 & 91 \nl
$\Lambda_v$ & velocity resolution & 65 \kms & \nodata & 60 \kms & \nodata \nl
\nl

\multicolumn{2}{l}{Observations} \nl
\hline 
$\delta\lambda$ & sampling increment & 0.96 \AA & 0.99 \AA & \nodata & 0.67 \AA\nl
$\delta v$ & sampling increment & 44 \kms & 45 \kms & \nodata & 40.5 \kms \nl
\nl
\enddata
\tablenotetext{a}{Values include 2$\times$2 binning.}
\end{deluxetable}

\clearpage
\begin{deluxetable}{lccc}
\tablecaption{Parameters of the two-cone Monte-Carlo simulation which
best reproduces the observed kinematics of the southern outflow in
M82.
\label{conetab}}
\small
\tablewidth{0pt}
\tablehead{ \colhead{} & \colhead{} & \colhead{Inner cone} & \colhead{Outer cone}}
\startdata
inner height      & $z_1$ [pc]           & 0 & 350 \nl
outer height      & $z_2$ [pc]           & 350 & 800 \nl
minimum diameter  & $D_1$ [pc]           & 375 & 390 \nl
maximum diameter  & $D_2$ [pc]           & 400 & 590 \nl
opening angle     & $\alpha$             & 5\arcdeg & 25\arcdeg \nl
inclination angle & $\iota$              & 5\arcdeg  & 15\arcdeg \nl
position angle    & $pa$                 & 150\arcdeg--165\arcdeg & 165\arcdeg \nl
radial velocity   & $v(r)$ [\kms] ([pc]) & \multicolumn{2}{c}{$525 + 0.13 r$} \nl
\enddata
\end{deluxetable}

% 
% Figures
%

% \clearpage
% \begin{figure}
% \epsscale{0.95}
% \plotone{fig01.eps}
% \end{figure}
% 
% \clearpage
% \begin{figure}
% \epsscale{0.472}
% \plotone{fig02.eps}
% \end{figure}
% 
% \clearpage
% \begin{figure}
% \epsscale{0.536}
% \plotone{fig03.eps}
% \end{figure}
% 
% \clearpage
% \begin{figure}
% \epsscale{0.472}
% \plotone{fig04.eps}
% \end{figure}
% 
% \clearpage
% \begin{figure}
% \epsscale{0.470}
% \plotone{fig05.eps}
% \end{figure}
% 
% \clearpage
% \begin{figure}
% \epsscale{0.95}
% \plotone{fig06.eps}
% \end{figure}
% 
% \clearpage
% \begin{figure}
% \epsscale{0.95}
% \plotone{fig07.eps}
% \end{figure}
% 
% \clearpage
% \begin{figure}
% \epsscale{0.95}
% \plotone{fig08.eps}
% \end{figure}
% 
% \clearpage
% \begin{figure}
% \epsscale{0.95}
% \plotone{fig09.eps}
% \end{figure}
% 
% \clearpage
% \begin{figure}
% \epsscale{0.95}
% \plotone{fig10.eps}
% \end{figure}
% 
% \clearpage
% \begin{figure}
% \epsscale{0.95}
% \plotone{fig11.eps}
% \end{figure}
% 
% \clearpage
% \begin{figure}
% \epsscale{0.60}
% \plotone{fig12.eps}
% \end{figure}
% 
% \clearpage
% \begin{figure}
% \epsscale{0.472}
% \plotone{fig13.eps}
% \end{figure}
% 
% \clearpage
% \begin{figure}
% \epsscale{0.95}
% \plotone{fig14.eps}
% \end{figure}
% 
% \clearpage
% \begin{figure}
% \epsscale{0.95}
% \plotone{fig15.eps}
% \end{figure}
% 
% \clearpage
% \begin{figure}
% \epsscale{0.95}
% \plotone{fig16.eps}
% \end{figure}


\newpage
\begin{thebibliography}{}
\bibitem[Achtermann \& Lacy 1995]{AL95} Achtermann, J. M., \& Lacy, J.
  H. 1995, \apj, 439, 163
  
\bibitem[Arav \& Li 1994]{AL94} Arav, N., \& Li, Z.-Y. 1994, \apj, 427,
  700
  
\bibitem[Arav, Li, \& Begelman 1994]{ALB94} Arav, N., Li, Z.-Y., \&
  Begelman, M.  C.  1994, \apj, 432, 62
  
\bibitem[Armus, Heckman, \& Miley 1990]{AHM90} Armus, L., Heckman, T.
  M., \& Miley, G. K. 1990, \apj, 364, 471
  
\bibitem[Armus \etal\ 1995]{AHWL95} Armus, L., Heckman, T.  M., Weaver,
  K. A., \& Lehnert, M. D. 1995, \apj, 445, 666
  
\bibitem[Axon \& Taylor 1978]{AT78} Axon, D. J., \& Taylor, K.  1978,
  \nat, 274, 37

\bibitem[Baum \etal\ 1993]{BODDP93} Baum, S. A., O'Dea, C. P.,
  Dallacassa, D., De Bruyn, A. G., \& Pedlar, A. 1993, \apj, 419, 553
  
\bibitem[Beck \etal\ 1978]{BLBT78} Beck, S. C., Lacy, J. H., Baas, F.,
  \& Townes, C. H. 1978, \apj, 226, 545

\bibitem[Becklin, Gatley, \& Werner 1982]{BGW82} Becklin, E. E.,
  Gatley, I., \& Werner, M. W. 1982, \apj, 258, 135
  
\bibitem[Bettoni \& Galletta 1982]{BG82} Bettoni, D., \& Galletta, G.
  1982, \aap, 113, 344
  
\bibitem[Bland \& Tully 1988]{BT88} Bland, J., \& Tully, R. B.  1988,
  \nat, 334, 43
  
\bibitem[Bland \& Tully 1989]{BT89} Bland, J., \& Tully, R. B. 1989,
  \aj, 98, 723
  
\bibitem[Bland-Hawthorn, Sokolowski, \& Cecil 1991]{BSC91}
  Bland-Hawthorn, J., Sokolowski, J., \& Cecil, G. 1991, \apj, 375, 78
  
\bibitem[Bland-Hawthorn \etal\ 1997]{BSVJ97} Bland-Hawthorn, J.,
  Shopbell, P. L., Veilleux, S., \& Jones, D. H. 1997, in preparation
  
\bibitem[Blecha \etal\ 1990]{BGHRB90} Blecha, A., Golay, M., Huguenin,
  D., Reichen, D., \& Bersier, D. 1990, \aap, 233, L9

\bibitem[Bloemen 1990]{B90} Bloemen, H., The Interstellar Disk-Halo
  Connection in Galaxies, IAU Symp.~144 (Kluwer Academic Publishers:
  Dordrecht)

\bibitem[Boer, Schulz, \& Keel 1992]{BSK92} Boer, B., Schulz, H., \&
  Keel, W. C. 1992, \aap, 260, 67

\bibitem[Bregman, Schulman, \& Tomisaka 1995]{BST95} Bregman, J. N.,
  Schulman, E., \& Tomisaka, K. 1995, \apj, 439, 155
  
\bibitem[Burbidge, Burbidge, \& Rubin 1964]{BBR64} Burbidge, E. M.,
  Burbidge, G. R., \& Rubin, V. C. 1964, \apj, 140, 942

\bibitem[Calzetti, Kinney, \& Storchi-Bergmann 1994]{CKS94} Calzetti,
  D., Kinney, A. L., \& Storchi-Bergmann, T. 1994, \apj, 429, 582

\bibitem[Calzetti \etal\ 1995]{CBGWB95} Calzetti, D., Bohlin, R. C.,
Gordon, K. D., Witt, A. N., \& Bianchi, L. 1995, \apjl, 446, L97
  
\bibitem[Castles, McKeith, \& Greve 1991]{CMG91} Castles, J., McKeith,
  C. D., \& Greve, A. 1991, Vistas~in~Astr., 34, 187
  
\bibitem[Chevalier \& Clegg 1985]{CC85} Chevalier, R. A., \& Clegg, A.
  W. 1985, \nat, 317, 44
  
\bibitem[Christopoulou \etal\ 1997]{CHSMTGMP97} Christopoulou, P. E.,
  Holloway, A. J., Steffen, W., Mundell, C. G., Thean, A. H. C.,
  Goudis, C. D., Meaburn, J., \& Pedlar, A. 1997, \mnras, 284, 385

\bibitem[Colbert \etal\ 1996a]{CBGOLTMC96} Colbert, E. J. M., Baum,
  S. A., Gallimore, J. F., O'Dea, C. P., Lehnert, M. D., Tsvetanov,
  Z. I., Mulchaey, J. S., \& Caganoff, S. 1996, \apjs, 105, 75

\bibitem[Colbert \etal\ 1996b]{CBGOC96} Colbert, E. J. M., Baum,
  S. A., Gallimore, J. F., O'Dea, C. P., \& Christensen, J. A. 1996,
  \apj, 467, 551
  
\bibitem[Colina \& P\'erez-Olea 1992]{CP92} Colina, L., \&
  P\'erez-Olea, D. E. 1992, \mnras, 259, 709

\bibitem[Cottrell 1977]{C77} Cottrell, G. A. 1997, \mnras, 178, 577
  
\bibitem[Courvoisier \etal\ 1990]{CRBGH90} Courvoisier, T. J.-L.,
  Reichen, M., Blecha, A., Golay, M., \& Huguenin, D. 1990, \aap, 238,
  63
  
\bibitem[Crawford \& Fabian 1992]{CF92} Crawford, C. S., \& Fabian, A.
  C. 1992, \mnras, 259, 265

\bibitem[Dahlem \etal\ 1997]{DPLHE97} Dahlem, M., Petr, M. G.,
  Lehnert, M. D., Heckman, T. M., \& Ehle, M. 1997, \aap, 320, 731
  
\bibitem[Davies 1974]{D74} Davies, R. D. 1974, in The Formation and
  Dynamics of Galaxies, ed. J. R. Shakeshaft, IAU Symp.~58 (Kluwer
  Academic Publishers: Dordrecht), p.~119
  
\bibitem[de Vaucouleurs \etal\ 1991]{ddCBPF91} de Vaucouleurs, G., de
  Vaucouleurs, A., Corwin, Jr., H.  G., Buta, R. J., Paturel, G., \&
  Fouqu\'e, P.  1991, Third Reference Catalogue of Bright Galaxies
  (Springer-Verlag: New York)
  
\bibitem[Doane 1993]{D93} Doane, J. S. 1993, PhD Thesis, University of
  California, Santa Cruz, California
  
\bibitem[Doane \& Mathews 1993]{DM93} Doane, J. S., \& Mathews, W. G.
  1993, \apj, 419, 573
  
\bibitem[Donahue \& Voit 1991]{DV91} Donahue, M., \& Voit, G. M. 1991,
  \apj, 381, 361

\bibitem[Dopita 1997]{D97} Dopita, M. A. 1997, private communication

\bibitem[Dopita, Binette, \& Schwartz 1982]{DBS82} Dopita, M. A.,
  Binette, L., \& Schwartz, R. D. 1982, \apj, 261, 183
  
\bibitem[Dopita \& Sutherland 1995]{DS95} Dopita, M. A., \& Sutherland,
  R. S. 1995, \apj, 455, 468
  
\bibitem[Dopita \& Sutherland 1996]{DS96} Dopita, M. A., \& Sutherland,
  R. S. 1996, \apjs, 102, 161

\bibitem[Duffy \etal\ 1987]{DEHH87} Duffy, P. B., Erickson, E. F.,
Haas, M. R., \& Houck, J. R. 1987, \apj, 315, 68
  
\bibitem[Elvius 1969]{E69} Elvius, A. 1969, Lick Obs.\ Bull., 7, 117
  
\bibitem[Elvius 1972]{E72} Elvius, A. 1972, \aap, 19, 193

\bibitem[Evans \& Dopita 1985]{ED85} Evans, I. N., \& Dopita,
  M. A. 1985, \apjs, 58, 125

\bibitem[Fabbiano, Heckman, \& Keel 1990]{FHK90} Fabbiano, G.,
  Heckman, T., \& Keel, W. C. 1990, \apj, 355, 442
  
\bibitem[Fabbiano \& Trinchieri 1984]{FT84} Fabbiano, G., \&
  Trinchieri, G. 1984, \apj, 286, 491
  
\bibitem[Fabbiano 1988]{F88} Fabbiano, G. 1988, \apj, 330, 672
  
\bibitem[Ferrara \etal\ 1996]{FBDG96} Ferrara, A., Bianchi, S.,
  Dettmar, R.-J., \& Giovanardi, C. 1996, \apj, 467, L69

\bibitem[Ferrara 1997]{F97} Ferrara, A. 1997, in The Local Bubble and
  Beyond, IAU Symp.~166 (Springer-Verlag: New York), preprint

\bibitem[Forbes, Boisson, \& Ward 1992]{FBW92} Forbes, D. A., Boisson,
  C., \& Ward, M. J. 1992, \mnras, 259, 293

\bibitem[Ford \etal\ 1985]{FCJLV85} Ford, H. C., Crane, P. C., Jacoby,
G. H., Lawrie, D. G., \& van der Hulst, J. M. 1985, \apj, 293, 132
  
\bibitem[Freedman \etal\ 1994]{FHMMLSKTFFGHHHI94} Freedman, W. L.,
  Hughes, S. M., Madore, B. F., Mould, J. R., Lee, M. G., Stetson, P.,
  Kennicutt, R. C., Turner, A., Ferrarese, L., Ford, H., Graham, J.
  A., Hill, R., Hoessel, J. G., Huchra, J., \& Illingworth, W. 1994,
  \apj, 427, 628

\bibitem[Gaffney \& Lester 1992]{GL92} Gaffney, N. I., \& Lester,
  D. F. 1992, \apj, 394, 139

\bibitem[Goad \& Gallagher 1985]{GG85} Goad, J. W., \& Gallagher,
  J. S. 1985, \apj, 297, 98

\bibitem[Hamilton \& Keel 1987]{HK87} Hamilton, D., \& Keel,
  W. C. 1987, \apj, 321, 211

\bibitem[Hayes 1970]{H70} Hayes, D. S. 1970, \apj, 159, 165
  
\bibitem[Hayes \& Latham 1975]{HL75} Hayes, D. S., \& Latham, D.  W.
  1975, \apj, 197, 593
  
\bibitem[Heckathorn 1972]{H72} Heckathorn, H. M. 1972, \apj, 173, 501
  
\bibitem[Heckman, Armus, \& Miley 1990]{HAM90} Heckman, T. M., Armus,
  L., \& Miley, G. K. 1990, \apjs, 74, 833
  
\bibitem[Heckman \etal\ 1995]{HDLFGW95} Heckman, T. M., Dahlem, M.,
  Lehnert, M. D., Fabbiano, G. A., Gilmore, D., \& Waller, W. H. 1995,
  \apj, 448, 98
  
\bibitem[Hennessy 1993]{H93} Hennessy, G. S. 1993, \baas, 25, 1328

\bibitem[Hennessy 1996]{He96} Hennessy, G. S. 1996, PhD Thesis,
  University of Virginia, Charlottesville, Virginia
  
\bibitem[Houck \etal\ 1984]{HSGH84} Houck, J. R., Shure, M. A., Gull,
  G. E., \& Herter, T. 1984, \apjl, 287, L11
  
\bibitem[Huang \etal\ 1994]{HTCCY94} Huang, Z. P., Thuan, T.  X.,
  Chevalier, R. A., Condon, J. J., \& Yin, Q. F. 1994, \apj, 424, 114
  
\bibitem[Hunter \& Gallagher 1990]{HG90} Hunter, D. A., \& Gallagher
  III, J. S. 1990, \apj, 362, 480
  
\bibitem[Ichikawa \etal\ 1994]{IvASTY94} Ichikawa, T., van Driel, W.,
  Aoki, T., Soyano, T., Tarusawa, K., \& Yoshida, S. 1994, \apj, 433,
  645

\bibitem[Jones, Tielens, \& Hollenbach 1996]{JTH96} Jones, A. P.,
  Tielens, A. G. G. M., \& Hollenbach, D. J. 1996, \apj, 469, 740

\bibitem[Keel 1983]{K83} Keel, W. 1983, \apj, 269, 466

\bibitem[Kobulnicky \& Skillman 1996]{KS96} Kobulnicky, H. A., \&
  Skillman, E. D. 1996, \apj, 471, 211
  
\bibitem[Koo \& McKee 1992a]{KM92a} Koo, B.-C., \& McKee, C. F.  1992,
  \apj, 388, 93
  
\bibitem[Koo \& McKee 1992b]{KM92b} Koo, B.-C., \& McKee, C. F.  1992,
  \apj, 388, 103

\bibitem[Krabbe \etal\ 1997]{KCTK97} Krabbe, A., Colina, L., Thatte,
N., \& Kroker, H. 1997, \apj, 476, 98
  
\bibitem[Kronberg, Biermann, \& Schwab 1985]{KBS85} Kronberg, P. P.,
  Biermann, P., \& Schwab, F. R. 1985, \apj, 291, 693

\bibitem[Lehnert \& Heckman 1995]{LH95} Lehnert, M. D., \& Heckman,
  T. M. 1995, \apjs, 97, 89

\bibitem[Lehnert \& Heckman 1996]{LH96} Lehnert, M. D., \& Heckman,
  T. M. 1996, \apj, 462, 651

\bibitem[L\'ipari, Colina, \& Macchetto 1994]{LCM94} L\'ipari, S.,
  Colina, L., \& Macchetto, F. 1994, \apj, 427, 174

\bibitem[Loiseau \etal\ 1990]{LNSWRK90} Loiseau, N., Nakai, N., Sofue,
  Y., Wielebinski, R., Reuter, H.-P., \& Klein, U. 1990, \aap, 228,
  331
  
\bibitem[Lynds \& Sandage 1963]{LS63} Lynds, C. R., \& Sandage, A. R.
  1963, \apj, 137 ,1005
  
\bibitem[Maran \etal\ 1991]{MOLNRSSS91} Maran, S. P., O'Connell, R.
  W., Landsman, W. B., Neff, S. G., Roberts, M. S., Smith, A. M.,
  Smith, E. P., \& Stecher, T. P. 1993, \baas, 23, 950
  
\bibitem[Mathews \& Baker 1971]{MB71} Mathews, W. G., \& Baker, J. C.
  1971, \apj, 170, 241
  
\bibitem[Mathis, Rumpl, \& Nordsieck 1977]{MRN77} Mathis, J.  S.,
  Rumpl, W., \& Nordsieck, K. H. 1977, \apj, 217, 425
  
\bibitem[McCarthy, Heckman, \& van Breugel 1987]{MHv87} McCarthy, P.
  J., Heckman, T. M., \& van Breugel, W. 1987, \aj, 93, 264
  
\bibitem[McKeith \etal\ 1993]{MCGD93} McKeith, C. D., Castles, J.,
  Greve, A., \& Downes, D. 1993, \aap, 272, 98
  
\bibitem[McKeith \etal\ 1995]{MGDP95} McKeith, C. D., Greve, A.,
  Downes, D., \& Prada, F. 1995, \aap, 293, 703
  
\bibitem[McLeod \etal\ 1993]{MRRK93} McLeod, K. K., Rieke, G. H.,
  Rieke, M. J., \& Kelly, D. M. 1993, \apj, 412, 111
  
\bibitem[Meurer, Freeman, \& Dopita 1989]{MFD89} Meurer, G. R.,
  Freeman, K. C., \& Dopita, M. A. 1989, \apss, 156, 141
  
\bibitem[Mihos \& Hernquist 1994]{MH94} Mihos, J. C., \& Hernquist, L.
  1994, \apjl, 431, L9
  
\bibitem[Miyashiro 1982]{M82} Miyashiro, G. M. 1982, Zodiac User's
    Manual, 2nd ed.

\bibitem[Moran \& Lehnert 1997]{ML97} Moran, E. C., \& Lehnert,
  M. D. 1997, \apj, 478, 172
  
\bibitem[Muxlow \etal\ 1994]{MPWASd94} Muxlow, T. W. B., Pedlar, A.,
  Wilkinson, P. N., Axon, D. J., Sanders, E. M., \& de Bruyn, A. G.
  1994, \mnras, 266, 455
  
\bibitem[Nakai \etal\ 1987]{NHHSHS87} Nakai, N., Hayashi, M., Handa,
  T., Sofue, Y., Hasegawa, T., \& Sasaki, M. 1987, \pasj, 39, 685
  
\bibitem[O'Connell \etal\ 1995]{OGHC95} O'Connell, R. W., Gallagher
  III, J. S., Hunter, D. A., \& Colley, W. N. 1995, \apjl, 446, L1
  
\bibitem[O'Connell \& Mangano 1978]{OM78} O'Connell, R. W., \& Mangano,
  J. J. 1978, \apj, 221, 62
  
\bibitem[Osterbrock 1989]{O89} Osterbrock, D. E. 1989, Astrophysics of
  Gaseous Nebulae and Active Galactic Nuclei (University Science
  Books: Mill Valley, California)
  
\bibitem[Osterbrock \& Martel 1992]{OM92} Osterbrock, D. E., \& Martel,
  A. 1992, \pasp, 104, 76
  
\bibitem[Ostriker \& Silk 1973]{OS73} Ostriker, J. P., \& Silk, J.
  1973, \apjl, 184, L113
  
\bibitem[P\'erez-Olea \& Colina 1996]{PC96} P\'erez-Olea, D. E., \&
  Colina, L. 1996, \apj, 468, 191

\bibitem[Pfeffermann \etal\ 1987]{PBHKMPRSZC87} Pfeffermann, E.,
  Briel, U. G., Hippmann, H., Kettenring, G., Metzner, G., Predehl,
  P., Reger, G., Stephan, K.-H., Zombeck, M. V., \& Chappell, J. 1987,
  SPIE Proceedings Vol.\ 733, 519
                    
\bibitem[Phillips 1993]{P93} Phillips, A. C. 1993, \aj, 105, 486
  
\bibitem[Phillips \etal\ 1983]{PBAC83} Phillips, M. M., Baldwin, J.
  A., Atwood, B., \& Carswell, R. F. 1983, \apj, 274, 558
  
\bibitem[Rice \etal\ 1988]{RLSNKLdH88} Rice, W., Lonsdale, C. J.,
  Soifer, B. T., Neugebauer, G., Kopan, E. L., Lloyd, L. A., de Jong,
  T., \& Habing, H. J. 1988, \apjs, 68, 91
  
\bibitem[Rieke \etal\ 1980]{RLTLT80} Rieke, G. H., Lebofsky, M. J.,
  Thompson, R. I., Low, F. J., \& Tokunaga, A. T. 1980, \apj, 238, 24
  
\bibitem[Rieke \etal\ 1993]{RLRT93} Rieke, G. H., Loken, K., Rieke, M.
  J., \& Tamblyn, P. 1993, \apj, 412, 99
  
\bibitem[Rodriguez \& Chaisson 1980]{RC80} Rodriguez, L. F., \&
  Chaisson, E. J. 1980, \apj, 238, 41
  
\bibitem[Rohan, Morrison, \& Sadun 1987]{RMS87} Rohan, M., Morrison,
  P., \& Sadun, A. 1987, in Star Formation in Galaxies, ed. C.
  J. Lonsdale, NASA Conf.~Pub. 2466, p.~485
  
\bibitem[Scarrott, Eaton, \& Axon 1991]{SEA91} Scarrott, S. M., Eaton,
  N., \& Axon, D. J. 1991, \mnras, 252, 12P
  
\bibitem[Schaaf \etal\ 1989]{SPBKS89} Schaaf, R., Pietsch, W.,
  Biermann, P. L., Kronberg, P. P., \& Schmutzler, T. 1989, \apj, 336,
  722
  
\bibitem[Schmidt, Angel, \& Cromwell 1976]{SAC76} Schmidt, G. D.,
  Angel, J. R. P., \& Cromwell, R. H. 1976, \apj, 206, 888
  
\bibitem[Seaquist, Bell, \& Bignell 1985]{SBB85} Seaquist, E.  R.,
  Bell, M. B., \& Bignell, R. C. 1985, \apj, 294, 546
  
\bibitem[Seaquist \& Odegard 1991]{SO91} Seaquist, E.  R., \& Odegard,
  N.  1991, \apj, 369, 320

\bibitem[Shapiro \& Benjamin 1991]{SB91} Shapiro, P. R., \& Benjamin,
  R. A. 1991, \pasp, 103, 923

\bibitem[Shen \& Lo 1995]{SL95} Shen, J., \& Lo, K. Y. 1995, \apjl,
  445, L99

\bibitem[Shopbell 1995]{S95} Shopbell, P. L. 1995, PhD Thesis, Rice
  University, Houston, Texas
  
\bibitem[Shopbell 1997]{S97} Shopbell, P. L. 1997, Zodiac+ User's
    Manual, 1st ed.
  
\bibitem[Shu \etal\ 1991]{SRLL91} Shu, F. H., Ruden, S. P., Lada, C.
  J., \& Lizano, S. 1991, \apjl, 370, L31
  
\bibitem[Slavin \& Cox 1993]{SC93} Slavin, J. D., \& Cox, D. P. 1993,
  \apj, 417, 187
  
\bibitem[Slavin, Shull, \& Begelman 1993]{SSB93} Slavin, J. D., Shull,
  J. M., \& Begelman, M. C. 1993, \apj, 407, 83
  
\bibitem[Smith 1993]{S93} Smith, S. J. 1993, \apj, 411, 570
  
\bibitem[Smith, Kennel, \& Coroniti 1993a]{SKC93a} Smith, S.  J.,
  Kennel, C. F., \& Coroniti, F. V. 1993, \apj, 411, 581
  
\bibitem[Smith, Kennel, \& Coroniti 1993b]{SKC93b} Smith, S.  J.,
  Kennel, C. F., \& Coroniti, F. V. 1993, \apj, 412, 82
  
\bibitem[Sofue \etal\ 1990]{SNHGRW90} Sofue, Y., Nakai, N., Handa, T.,
  Golla, G., Reuter, H.-P., \& Wielebinski, R. 1990, in The
  Interstellar Disk-Halo Connection in Galaxies, ed. H. Bloemen, IAU
  Symp.~144 (Kluwer Academic Publishers: Dordrecht), p.~9
  
\bibitem[Soifer, Houck, \& Neugebauer 1987]{SHN87} Soifer, B. T.,
  Houck, J. R., \& Neugebauer, G. 1987, \araa, 25, 187
  
\bibitem[Sokolowski 1992]{S92} Sokolowski, J. 1992, PhD Thesis, Rice
  University, Houston, Texas
  
\bibitem[Solinger 1969]{S69} Solinger, A. B. 1969, \apj, 155, 403
  
\bibitem[Solinger \& Markert 1975]{SM75} Solinger, A. B., \& Markert,
  T. 1975, \apj, 197, 309
  
\bibitem[Solinger, Morrison, \& Markert 1977]{SMM77} Solinger, A. B.,
  Morrison, P., \& Markert, T. 1977, \apj, 211, 707
  
\bibitem[Stark \& Carlson 1984]{SC84} Stark, A. A., \& Carlson, E. R.
  1984, \apj, 279, 122
  
\bibitem[Stecher \etal\ 1992]{Se92} Stecher, T. P., \etal\ 1992, \apj,
  395, L1
  
\bibitem[Strickland, Ponman, \& Stevens 1996]{SPS96} Strickland, D.
  K., Ponman, T. J., \& Stevens, I. R. 1996, \aap, 320, 378
  
\bibitem[Suchkov \etal\ 1994]{SBHL94} Suchkov, A. A., Balsara, D. S.,
  Heckman, T. M., \& Leitherer, C. 1994, \apj, 430, 511
  
\bibitem[Suchkov \etal\ 1996]{SBHB96} Suchkov, A. A., Berman, V. G.,
  Heckman, T. M., \& Balsara, D. S. 1996, \apj, 463, 528

\bibitem[Sulentic \& Arp 1987]{SA87} Sulentic, J. W., \& Arp,
  H. C. 1987, \apj, 319, 693
  
\bibitem[Sutherland, Bicknell, \& Dopita 1993]{SBD93} Sutherland,
  R. S., Bicknell, G. V., \& Dopita, M. A. 1993, \apj, 414, 510

\bibitem[Tammann \& Sandage 1968]{TS68} Tammann, G. A., \& Sandage, A.
  R. 1968, \apj, 151, 825
  
\bibitem[Telesco \etal\ 1991]{TCJDD91} Telesco, C. M., Campins, H.,
  Joy, M., Dietz, K., \& Decher, R. 1991, \apj, 369, 135

\bibitem[Telesco \& Gezari 1992]{TG92} Telesco, C. M., \& Gezari,
  D. Y. 1992, \apj, 395, 461

\bibitem[Tielens \etal\ 1994]{TMSH94} Tielens, A. G. G. M., McKee,
  C. F., Seab, C. G., \& Hollenbach, D. J. 1994, \apj, 431, 321
  
\bibitem[Tomisaka \& Bregman 1993]{TB93} Tomisaka, K., \& Bregman, J.
  N. 1993, \pasj, 45, 513
  
\bibitem[Tomisaka \& Ikeuchi 1988]{TI88} Tomisaka, K., \& Ikeuchi, S.
  1988, \apj, 330, 695
  
\bibitem[Tr\"umper 1984]{T84} Tr\"umper, J. 1984, Phys.\ Scripta, T7,
  209
  
\bibitem[Tsuru \etal\ 1990]{TOMMK90} Tsuru, T., Ohashi, T., Makishima,
  K., Mihara, T., \& Kondo, H. 1990, \pasj, 42, L75
  
\bibitem[Tsuru \etal\ 1994]{THAKFIIOPRN94} Tsuru, T., Hayashi, I.,
  Awaki, H., Koyama, K., Fukazawa, Y., Ishisaki, Y., Iwasawa, K.,
  Ohashi, T., Petre, R., Rasmussen, A., \& Nousek, J. 1994, in New
  Horizon of X-ray Astronomy - First Results from ASCA, eds. F. Makino
  and T. Ohashi (Universal Academy Press, Japan), p.~529
  
\bibitem[Veilleux \etal\ 1994]{VCBTFS94} Veilleux, S., Cecil, G.,
  Bland-Hawthorn, J., Tully, R. B., Filippenko, A. V., \& Sargent, W.
  L. W. 1994, \apj, 433, 48
  
\bibitem[Veilleux \& Osterbrock 1987]{VO87} Veilleux, S., \&
  Osterbrock, D. E. 1987, \apjs, 63, 295
  
\bibitem[Visvanathan \& Sandage 1972]{VS72} Visvanathan, N., \&
  Sandage, A. R. 1972, \apj, 176, 57
  
\bibitem[Visvanathan 1974]{V74} Visvanathan, N. 1974, \apj, 192, 319
  
\bibitem[Voit, Donahue, \& Slavin 1994]{VDS94} Voit, G. M., Donahue,
  M., \& Slavin, J. D. 1994, \apjs, 95, 87
  
\bibitem[Waller, Gurwell, \& Tamura 1992]{WGT92} Waller, W. H.,
  Gurwell, M., \& Tamura, M. 1992, \aj, 104, 63
  
\bibitem[Watson, Stanger, \& Griffiths 1984]{WSG84} Watson, M. G.,
  Stanger, V., \& Griffiths, R.  E.  1984, \apj, 286, 144
  
\bibitem[White \& Chevalier 1983]{WC83} White III, R. E., \& Chevalier,
  R. A. 1983, \apj, 275, 69
  
\bibitem[Williams, Caldwell, \& Schommer 1984]{WCS84} Williams, T. B.,
  Caldwell, N., \& Schommer, R. A. 1984, \apj, 281, 579
  
\bibitem[Yun, Ho, \& Lo 1994]{YHL94} Yun, M. S., Ho, P. T. P., \& Lo,
  K. Y. 1994, \nat, 372, 530

\end{thebibliography}
\end{document}